\documentclass[fleqn,usenatbib,usedcolumn]{mnras}

\usepackage{newtxtext,newtxmath}
\usepackage[T1]{fontenc}
\usepackage{ae,aecompl}

\usepackage{multirow}
\usepackage{caption}
\usepackage{dsfont}
\usepackage[capitalise]{cleveref}
\usepackage{xcolor}
\usepackage{arydshln}
\usepackage{commath}

\usepackage{graphicx}
\usepackage{bbold}
\usepackage{graphicx}	% Including figure files
\usepackage{amsmath}	% Advanced maths commands
\usepackage{amssymb}	% Extra maths symbols
\usepackage{csquotes}
\usepackage{anyfontsize}
\usepackage{physics}

\newcommand{\Gyr}{\,\hbox{Gyr}}
\newcommand{\Mpc}{\,\hbox{Mpc}}
\newcommand{\Msol}{\,{\hbox{M}_\odot}}
\newcommand{\noop}[1]{}

\title[How feedback shapes galaxies]{How feedback shapes galaxies: an analytic model}

\author[J. Salcido et al.]{Jaime Salcido,$^{\href{https://orcid.org/0000-0002-8918-5229}{\includegraphics[scale=0.04]{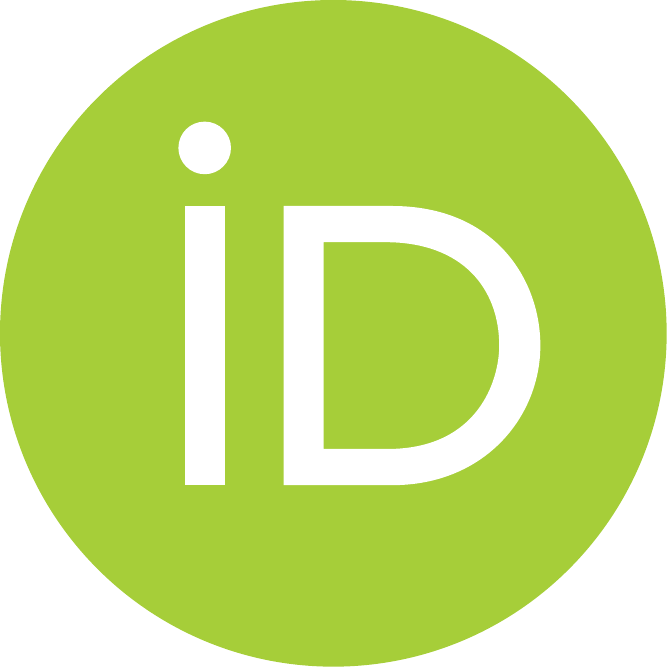}} 1, 2}$\thanks{E-mail:
\href{mailto:J.SalcidoNegrete@ljmu.ac.uk}{J.SalcidoNegrete@ljmu.ac.uk}} 
Richard G. Bower,$^{\href{https://orcid.org/0000-0002-5215-6010}{\includegraphics[scale=0.04]{ORCIDiD.pdf}}\, 2}$ 
Tom Theuns$^{\href{https://orcid.org/0000-0002-3790-9520}{\includegraphics[scale=0.04]{ORCIDiD.pdf}}\, 2}$ 
\\ \\
$^{1}$ Astrophysics Research Institute, Liverpool John Moores University, 146 Brownlow Hill, Liverpool L3 5RF, UK \\
$^{2}$ Institute for Computational Cosmology, Department of Physics, Durham University, South Road, Durham, DH1 3LE, UK}

\date{Accepted XXX. Received YYY; in original form ZZZ}

\pubyear{2019}

\begin{document}
\label{firstpage}
\pagerange{\pageref{firstpage}--\pageref{lastpage}}
\maketitle

\begin{abstract}
 We introduce a simple analytic model of galaxy formation that links the growth of dark matter haloes in a cosmological background to the build-up of stellar mass within them. The model aims to identify the physical processes that drive the galaxy-halo co-evolution through cosmic time. The model restricts the role of baryonic astrophysics to setting the relation between galaxies and their haloes. Using this approach, galaxy properties can be directly predicted from the growth of their host dark matter haloes. We explore models in which the effective star formation efficiency within haloes is a function of mass (or virial temperature) and independent of time. Despite its simplicity, the model reproduces self-consistently the shape and evolution of the cosmic star formation rate density, the specific star formation rate of galaxies, and the galaxy stellar mass function, both at the present time and at high redshifts. By systematically varying the effective star formation efficiency in the model, we explore the emergence of the characteristic shape of the galaxy stellar mass function. The origin of the observed double Schechter function at low redshifts is naturally explained by two efficiency regimes in the stellar to halo mass relation, namely, a stellar feedback regulated stage, and a supermassive black hole regulated stage. By providing a set of analytic differential equations, the model can be easily extended and inverted, allowing the roles and impact of astrophysics and cosmology to be explored and understood.
\end{abstract}


%%%%%%%%%%%%%%%%%%%%%%%%%%%%%%%%%%%%%%%%%%%%%%%%%%

%%%%%%%%%%%%%%%%% BODY OF PAPER %%%%%%%%%%%%%%%%%%

\section{Introduction}\label{sec:intro}

The co-evolution between galaxies and their haloes is perhaps one of the most fundamental aspects of every galaxy formation model. In the current paradigm of galaxy formation, every galaxy forms within a dark matter halo. However, understanding the relationship between a dark matter halo and the galaxies it hosts is not a trivial exercise due to our lack of detailed understanding of the complex baryonic process involved. 

In a standard Lambda Cold Dark Matter ($\Lambda$CDM) cosmology, gravitationally bound dark matter structures build up hierarchically, by a combination of the smooth accretion of surrounding matter and continuous merging with smaller structures \citep{white_rees78, qu_chronicle_2017}. The formation and evolution of galaxies within these haloes is thought to be a highly self-regulated process, in which galaxies tend to evolve towards a quasi-equilibrium state where the gas outflow rate balances the difference between the gas inflow rate and the rate at which gas is locked up in stars and black holes (BHs) \citep[e.g.][]{white_galaxy_1991,finlator_origin_2008,bouche_impact_2010,schaye_physics_2010,dave_analytic_2012,bower_dark_2017}. Consequently, galaxy formation is thought to be determined on the one hand by the formation and growth of dark matter haloes, which depends solely on the cosmological background, and on the other hand, by the regulation of the gas content in these haloes, which in turn depends on complex baryonic processes such as radiative cooling, stellar mass loss, and feedback from stars and accreting BHs. This co-evolution process results in a tight correlation between the properties of galaxies and their dark matter haloes (see e.g. \citealt{wechsler_connection_2018} for a review). 

A fundamental requirement for a successful galaxy formation model, is to  reproduce the relation between stellar mass and halo mass inferred from observations. However probing the dark matter distribution and its evolution represents an observational challenge. Direct observational probes include galaxy-galaxy lensing \citep[e.g.][]{brainerd_mass--light_2003,hoekstra_properties_2004,hudson_cfhtlens:_2015} and the kinematics of satellite galaxies \citep[e.g.][]{zaritsky_satellites_1993,van_den_bosch_probing_2004,norberg_massive_2008}. However, direct observation techniques are limited to low redshifts ($z<1$), due to the difficulty of resolving individual distant galaxies. Indirect methods include, for example, comparing the abundance and clustering properties of galaxy samples with predictions from phenomenological halo models \citep[e.g.][]{neyman_theory_1952,berlind_halo_2002,cooray_halo_2002,cowley_galaxyhalo_2018}. This method however, depends heavily on the underlying modelling and assumptions, for example, the bias with which haloes trace the underlying matter distribution. 

From the theoretical point of view, the formation and evolution of dark matter haloes is largely considered a ``solved problem'' (see however, \citealt{van_den_bosch_disruption_2018}). Using extremely accurate measurements of the density perturbations imprinted onto the cosmic microwave background radiation fluctuations as initial conditions \citep[e.g.][]{planck_collaboration_planck_2016}, many different groups have produced convergent results using large cosmological N-body simulations \citep[e.g][]{springel_simulations_2005, klypin_dark_2011,trujillo-gomez_galaxies_2011,angulo_scaling_2012,fosalba_mice_2015}.

On the other hand, the complex physics of galaxy formation still has many open questions. Different approaches have been used to model the intricate baryonic physics of galaxy formation. The most widely used technique combines the evolution of dark matter with either a \textit{semi-analytical} \citep[e.g.][]{cole_recipe_1994, somerville_semi-analytic_2008, henriques_galaxy_2015, lacey_unified_2016} or \textit{hydrodynamical} \citep[e.g.][]{vogelsberger_introducing_2014,schaye_eagle_2015,dave_mufasa:_2016,dubois_horizon-agn_2016,pillepich_simulating_2018} treatment of the baryonic processes involved. A key ingredient in both methods that has led us to a better understating of the physics of galaxy formation is the use of physically motivated models for feedback processes (see \citealt{somerville_physical_2015,naab_theoretical_2017} for a comprehensive review). 

An alternative approach known as \textit{empirical modelling} takes the advantage of the vast number of observational data sets from large galaxy surveys and relate statistical galaxy scaling relations to the evolution of dark matter haloes without assuming strong physical priors \citep[e.g.][]{behroozi_average_2013,moster_galactic_2013,rodriguez-puebla_is_2016,behroozi_universemachine:_2018, moster_emerge_2018, Grylls:2019}.

While all of these approaches have been very productive, the increasing complexity of the models and simulations make it difficult to pinpoint and understand the fundamental physics driving the results. For instance, in cosmological simulations, ``sub-grid'' physics are implemented as micro phenomena that depend only on local gas properties from which macroscopic patterns emerge. However, it is hard to track down the link between what emerges and why \citep[e.g.][]{bower_dark_2017}. In this paper, we examine this issue in detail by adopting the opposite approach. We develop a fully analytic model of galaxy formation derived from a simple relation between the star formation rate and halo growth rate that disentangles the role of cosmology from the role of astrophysics in the galaxy formation process. Our model restricts the role of baryonic astrophysics to setting the relation between galaxies and their haloes. With this simple relation, we can use an analytic approximation to the growth of dark matter haloes to predict galaxy properties. By providing a set of analytic equations, the model can be easily ``inverted'' and allows for rapid experiments to be conducted, providing a powerful tool to explore the differential effects of baryonic physics, averaged over galaxy scales. Despite its simplicity, the model reproduces self-consistently the shape and evolution of the cosmic star formation rate (SFR) density, the specific star formation rate (sSFR) of galaxies, and the galaxy stellar mass function (GSMF), both at the present time and at high redshift. 

We validate our results by comparing to numerical hydrodynamic simulations from the \textsc{eagle} project. The \textsc{eagle} simulation suite\footnote{The galaxy and halo catalogues of the simulation suite, as well as the particle data, are publicly available at \url{http://www.eaglesim.org/database.php} \citep{mcalpine_eagle_2016}.} \citep{schaye_eagle_2015,crain_eagle_2015} consists of a large number of cosmological hydrodynamical simulations that include different resolutions, simulated volumes and physical models. These simulations use advanced smoothed particle hydrodynamics (SPH) and state-of-the-art subgrid models to capture the unresolved physics. A complete description of the code and physical parameters used can be found in \citet{schaye_eagle_2015}. Here we compare to the \textsc{eagle} reference simulations that used a flat, $\Lambda$CDM cosmology with parameters ($\Omega_m=0.307$, $\Omega_\Lambda = 0.693$, $h=0.6777$, $\sigma_8 = 0.8288$, $n_s = 0.9611$) consistent with the \cite{planck_collaboration_planck_2014} results. The calibration strategy of the \textsc{eagle} simulations is described in detail by \citet{crain_eagle_2015}, who also presented additional simulations to demonstrate the effect of parameter variations.

The layout of this paper is as follows: In \cref{sec:model} we introduce the analytic model of galaxy formation. We present two models of the effective star formation efficiency: A time-independent efficiency which depends only on halo mass, and an efficiency that depends on the virial temperature of the halo. In \cref{sec:impact}, we explore the effect of the different efficiency parameters in the galaxy formation outputs. Namely, the cosmic star formation rate density, the specific star formation rate of galaxies, and the galaxy stellar mass function. In \cref{sec:fit} we compare the results from our model to different observational datasets. We also discuss the need for a time-evolving efficiency in order to reproduce the rapid evolution of the GSMF. We discuss the limitations of our model, and summarise our conclusions in \cref{sec:con}. 

\section{An analytic model of galaxy formation}\label{sec:model}

\subsection{The effective star formation efficiency}

The formation, evolution and abundance of dark matter haloes can be predicted accurately when the cosmology and dark matter model (i.e. cold, warm, self-interacting, etc.) is known. Although these processes are highly non-linear, the underlying physics is well understood. However, the gas and stellar content of haloes is much less well understood because of the intrinsic complexity of the baryonic processes, such as cooling, star formation and feedback, that drive it. An empirical approach to populating dark matter haloes with galaxies is to focus on the relation between stellar mass and halo mass inferred from observations. We write 
this relation as
\begin{equation}\label{eq:SHMR}
	\log_{10}\left(\frac{M_*}{10^{12}\Msol}\right)=\varepsilon(M_h, t) \log_{10}\left(\frac{M_h}{10^{12}\Msol}\right) + \log_{10}\mathcal{N}(t),
\end{equation}
where $M_*$ and $M_h$ are the central galaxy stellar mass and host halo mass respectively, $\mathcal{N}(t)$ is a normalisation factor, and $\varepsilon(M_h, t)$ is the logarithmic slope of the stellar to halo mass relation (SHMR). Allowing $\mathcal{N}(t)$ to be a random variable would account for the scatter in the relation, but in this paper we will focus on the mean relation
and replace $\mathcal{N}$ by its expectation value. $\varepsilon$ is closely related to the galaxy formation efficiency of haloes, and we will explore this connection further below. 

Probing the dark matter distribution and its evolution directly represents an observational challenge. Perhaps the simplest, and most commonly used alternative technique is to use (sub-) halo abundance matching to determine the typical SHMR \citep[e.g.][]{behroozi_average_2013, moster_galactic_2013}. In essence, the SHMR is derived by mapping the theoretical halo mass function and the observed abundance of galaxies given by the GSMF, 
\begin{equation}\label{eq:GSMF}
		\phi(M_*) \equiv \frac{\dd n_{\mathrm gal}}{\dd \mathrm{log}_{10}M_*}
		= \varepsilon ^{-1} \, \frac{\dd n_{\mathrm h}}{\dd \mathrm{log}_{10}M_h},
\end{equation}
where $n_{\mathrm gal}$ and $n_{\mathrm h}$ are the co-moving abundances of galaxies and haloes respectively. A subtlety here is that $M_*$ refers to the stellar mass of the central object in the halo. More complex formulations of the abundance matching method allow for the contribution of satellite galaxies to the mass function, but we will keep to the simple approach. This is adequate if the stellar mass function is dominated by central galaxies \citep[e.g.][]{yang_galaxy_2009,lan_galaxy_2016}.

Abundance matching studies have consistently shown that a simple picture of the galaxy population is consistent with much of the observational data. The SHMR is a strong function of halo mass but a weak function of cosmic time; it can be approximated well by two power laws that connect at a stellar mass that corresponds roughly to the knee of the stellar mass function \citep[e.g.][]{moster_constraints_2010, yang_evolution_2012, mitchell_evolution_2016}. We use this as the basis of a simple model that couples the build-up of dark matter haloes and the build-up of galaxy stellar mass.

We must be careful, however, to distinguish the instantaneous efficiency with which infalling baryons are converted into stars,
\begin{equation}\label{eq:eps_star}
    \epsilon_* = \frac{\dot{M}_*}{f_b \dot{M}_h}
\end{equation}
(where $f_b = {\Omega_b}/{\Omega_m}$ is the cosmic baryon fraction)
and the integral of this growth over the history of the halo, \cref{eq:SHMR}.
Note that $\dot{M}_*$ includes both star formation within the central object, and the accretion of infalling stars. We will need to distinguish between the two in order to relate the stellar mass growth to the observed star formation rate density.

In Appendix A, we show that the build up of dark matter haloes can be described
analytically using recent developments of the linear theory original described by \cite{press_formation_1974}. This allows the abundance and growth rates of haloes to be derived from the power spectrum of density fluctuations in the early universe. \Cref{eq:GSMF} provides a promising approach to connect the growth of haloes to the formation of galaxies, and the observational results suggest that a good starting point is to consider a time-independent SHMR that depends only on halo mass. In addition, we will examine a model in which the efficiency of star formation depends on the halo's virial velocity, as it reflects the evolution of the gravitational potential of the halo \citep[e.g.][]{sharma_2019}.

\begin{figure}
\centering 
\includegraphics[width=0.48\textwidth]{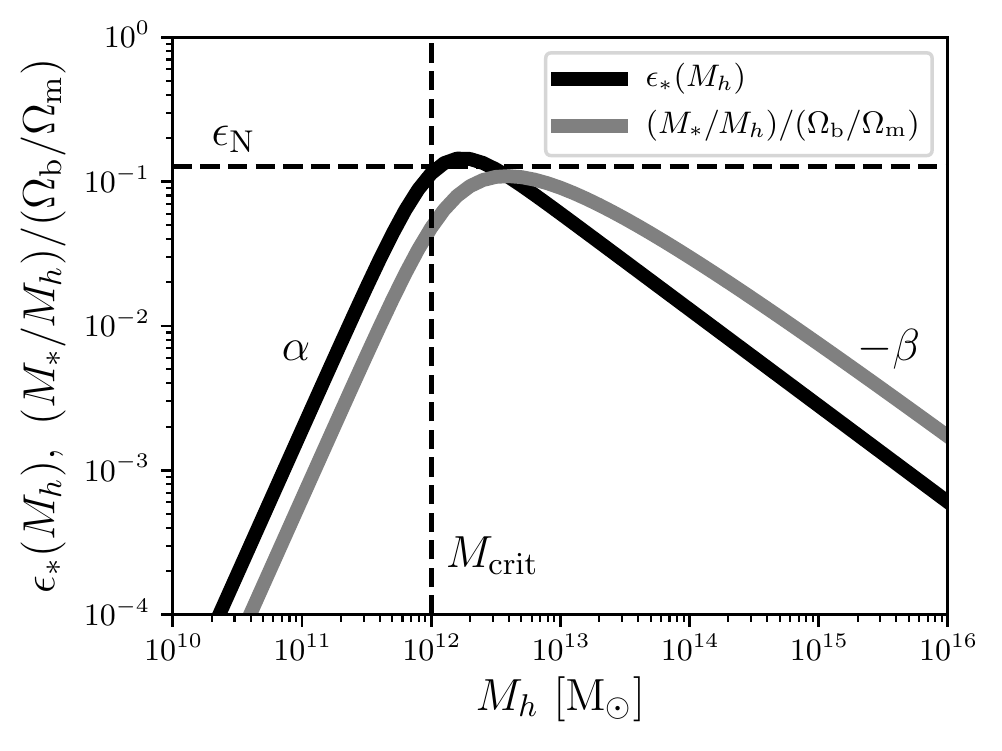} 
 \vspace{-1.5em}
 \caption{Parametrisation of the effective star formation efficiency $\epsilon_*$ provided in \cref{eq:epsilon}. $\epsilon_\mathrm{N}$ is the normalisation parameter, $\alpha$ and $\beta$ determine the slope of the efficiency at low and high masses respectively, and $M_\mathrm{crit}$ locates the transition mass, or peak efficiency. The SHMR is shown for comparison. $\epsilon_*$ has the same slopes as $M_*/M_h$, i.e. $\alpha$ and $\beta$, but the normalisation of $M_*/M_h$ is different by a factor of $1 / (1 + \alpha)$, and $1 / (1 - \beta)$ for low and high mass haloes respectively. }
 \label{fig:epsilon}
\end{figure}

\begin{figure*}
\centering 
\includegraphics[width=.95\textwidth]{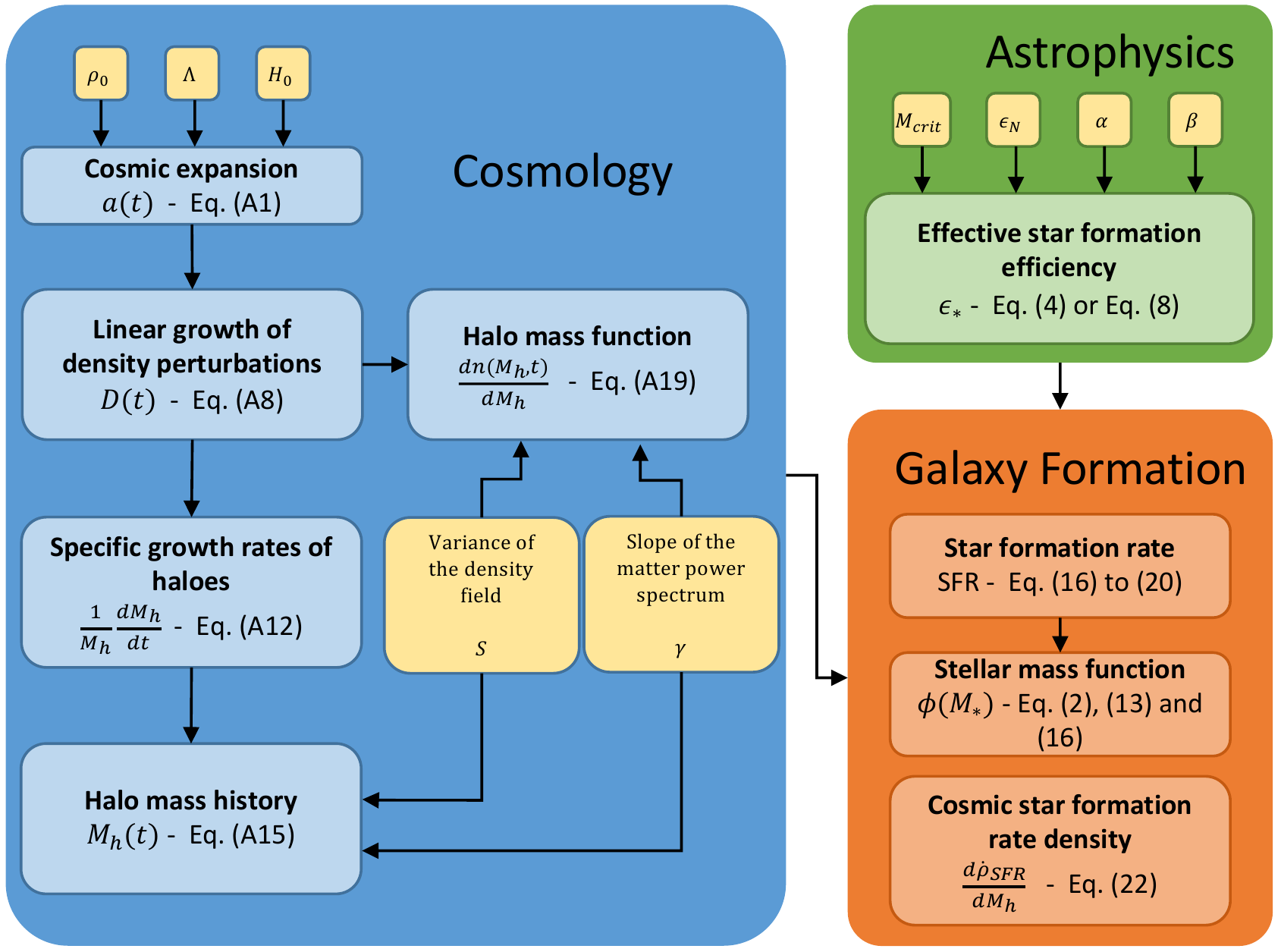}
 \caption{A schematic diagram of the analytic model of galaxy formation. All components in the blue block depend solely on cosmology. By using the Taylor expansion solution to the Friedmann equations in \citet{salcido_impact_2018}, all the cosmological components can be calculated analytically for a given cosmology defined by the parameters $\rho_0$, $\Lambda$, $H_0$, and the shape of the matter power-spectrum parametrised by $S$ and $\gamma$. All astrophysical processes (green) enter into the model in terms of the effective star formation efficiency $\epsilon_*$, which is fully described by the four free parameters $M_\mathrm{crit}$, $\epsilon_\mathrm{N}$, $\alpha$ and $\beta$ in \cref{eq:epsilon}. The galaxy formation outputs are summarised in the orange block.}
 \label{fig:schematic}
\end{figure*}

\begin{itemize}
	\item  \vspace{.15cm} \textit{Halo mass-dependent efficiency (Model I)} \vspace{.15cm} \label{sec:model1}

In order to model $\epsilon_*$ in \cref{eq:eps_star}, we begin by assuming that  the efficiency of conversion of infalling baryons into stars depends only on halo mass \citep[e.g.][]{rodriguez-puebla_is_2016, salcido_impact_2018, tacchella_redshift-independent_2018}. Motivated by the results from abundance matching techniques \citep[e.g.][]{behroozi_average_2013, rodriguez-puebla_is_2016}, that estimate a galaxy formation efficiency that peaks at masses similar to Milky-Way sized haloes (${\sim} 10^{12} \Msol$) and falls steeply for higher and lower masses, we model $\epsilon_*$ as a double power law, a similar parametrisation as that proposed by \cite{moster_constraints_2010}:
\begin{equation}\label{eq:epsilon}
\epsilon_*(M_h) = 2 \epsilon_\mathrm{N} \left[ \left(\frac{M_h}{M_\mathrm{crit}}\right)^{-\alpha} + \left(\frac{M_h}{M_\mathrm{crit}}\right)^{\beta} \right]^{-1},
\end{equation}
where $\epsilon_\mathrm{N}$ is a normalisation parameter, and $\alpha$ and $-\beta$ are the power-law slopes at low and high mass respectively. The maximum efficiency occurs at halo mass $M_\mathrm{crit}$. To agree with observational data, the values of $\alpha$ and $\beta$ are typically positive, i.e. at low masses, star formation is suppressed because of the efficiency of feedback from star formation, and at higher masses, the cooling of the inflowing gas is suppressed by heating from BHs \citep[e.g.][]{white_galaxy_1991,bower_breaking_2006,benson_g_2012,haas_physical_2013}.

Because of our assumption that $\epsilon_*$ depends only on $M_h$, we can integrate to determine $M_*$ without needing to know the time evolution of the
halo mass.
\begin{equation}\label{eq:mstar}
\begin{aligned}
M_* = & \int_0^{M_h} \epsilon_* f_b dM'_h \\
   = & \frac{2\epsilon_\mathrm{N}}{1 + \alpha} f_b M_\mathrm{crit} 
      \left(\frac{M_h}{M_\mathrm{crit}}\right)^{1+\alpha}
      F(\eta, \mathbb{z})
\end{aligned}
\end{equation}
where $\eta=(1+\alpha)/(\alpha+\beta)$, $\mathbb{z}=\left({M_h}/{M_\mathrm{crit}}\right)^{\alpha+\beta}$ and
\begin{equation}\label{eq:f_integral}
\begin{aligned}
F(\eta,\mathbb{z}) = & {\eta} \int_0^1 \frac{x^{\eta -1}} {(1+\mathbb{z} x)} dx \\
   = & _2F_1(1, \eta; 1+\eta; -\mathbb{z}), 
\end{aligned}
\end{equation}
where\footnote{We have used the symbol $\mathbb{z}$ to differentiate from redshift $z$.} $_2F_1(a,b;c;\mathbb{z})$, is the Gaussian hypergeometric function. For values of $\alpha > 0$ and $0 < \beta < 1$, in the limit $M_h{\ll}M_\mathrm{crit}$, $\lim_{\mathbb{z} \to 0} F(\eta,\mathbb{z}) = 1$, while for $M_h{\gg}M_\mathrm{crit}$, $F(\eta,\mathbb{z})\approx 1/((\eta-1) \,\mathbb{z})$. Differentiating \cref{eq:mstar}, the logarithmic slope of the stellar mass halo mass relation, $\varepsilon$, can be written in a simple analytic form,
\begin{equation}\label{eq:logslope}
\varepsilon = \frac{\mathrm{dlog} M_*}{\mathrm{dlog} M_h} = \frac{1 + \alpha}{(1+\mathbb{z})F(\eta,\mathbb{z})}.
\end{equation}
\Cref{eq:logslope} describes a smooth transition in slope from $(1+\alpha)$ for $M_{h}{\ll}M_{\mathrm{crit}}$, to $(1-\beta)$ for $M_{h}{\gg}M_{\mathrm{crit}}$. 

\Cref{fig:epsilon} shows an illustration of the effective star formation efficiency $\epsilon_*$ as a function of $M_h$. The parametrisation provides a smooth a transition from the $\alpha$ dominated regime (for low halo masses), to the $\beta$ dominated regime (for high halo masses). The figure shows that $\epsilon_*$ has the same slopes as $M_*/M_h$, i.e. $\alpha$ and $\beta$, but the normalisation of $M_*/M_h$ is different by a factor of $1 / (1 + \alpha)$, and $1 / (1 - \beta)$ for low and high mass haloes respectively. We note that for our chosen parametrisation, $M_*/M_h$ is closely approximated by a double power law. 

\item \vspace{.15cm} \textit{Virial temperature-dependent efficiency (Model II)} \vspace{.15cm}

As we will discuss in \cref{sec:impact}, an efficiency dependent only on halo mass turns out to be a very good approximation of the stellar mass build up of galaxies because most of the stellar mass builds up when the mass of the halo has roughly its current value. However, a time-independent efficiency model significantly under predicts the abundance galaxies at high redshifts ($z>4$), which hints at the need for a time-evolving efficiency model. A purely empirical approach \citep[e.g][]{moster_emerge_2018, behroozi_universemachine:_2018} would relax the physical priors and let, in this case, the four efficiency parameters in the model to evolve freely in time. Instead, we consider an alternative model in which $\epsilon_*$ depends only on the virial temperature of the halo $T_\mathrm{vir}$ (and hence the gravitational potential of the halo). 

Considering the energetics of galaxy winds suggests that, wind that marginally escape the gravitational binding energy of the galaxy's halo can carry a higher mass loading in lower mass haloes \citep{dekel_silk_86, white_galaxy_1991, dave_11, sharma_2019}. Since the energy required for escape depends on the halo virial temperature, $T_\mathrm{vir}$, this leads to an inverse scaling of the mass loading, and hence, star formation efficiency. At sufficiently high mass, the energy associated with individual supernovae becomes smaller than the halo binding energy. This may lead to the accumulation of gas around the central BH, and consequently a wind driven by BH accretion, rather than star formation \citep{dubois_2015}. A related argument can also be made based on the buoyancy of gas heated by star formation. \cite{bower_dark_2017} discusses a possible physical origin of a transition from where star formation driven outflows get hotter than the virial temperature of the halo and can escape (i.e. supernovae energy, or entropy, is much greater than the halo binding energy), to where outflows stall inside the halo triggering star formation and BH growth. 

In order to explore these effects, we model the effective star formation efficiency as a function of the halo's virial temperature using the same double power law parametrisation as in \cref{eq:epsilon}, 
\begin{equation}\label{eq:epsilon_T_vir}
\epsilon_*(T_\mathrm{vir}) = 2 \epsilon_\mathrm{N} \left[ \left(\frac{T_\mathrm{vir}}{T_\mathrm{crit}}\right)^{-\alpha} + \left(\frac{T_\mathrm{vir}}{T_\mathrm{crit}}\right)^{\beta} \right]^{-1},
\end{equation}
Note that the relation between halo mass and virial temperature depends on redshift (see \cref{eq:M200,eq:T_vir}), as the density of collapsed haloes decrease as the universe expands. As a result, it is not possible to analytically determine $M_*(M_h,t)$, but the required integrals can easily be evaluated numerically. We discuss this model further in \cref{sec:t_vir}.

\end{itemize}

\subsection{Halo definition}

Dark matter haloes are typically identified by growing a sphere outwards from the potential minimum of the dark matter halo out to a radius where the mean interior density equals a fixed multiple of the critical or mean density of the Universe, causing an artificial `pseudo-evolution' of dark matter haloes by changing the radius of the halo \citep{diemer_pseudo-evolution_2013}. Star formation, however, is governed by the amount of gas that enters these haloes and reaches their central regions. \cite{wetzel_physical_2015} show that the growth of dark matter haloes is subject to this `pseudo-evolution', whereas the accretion of gas is not. Because gas is able to cool radiatively, it decouples from the dark matter, tracking the accretion rate near a radius of $R_{200\mathrm{m}}$, the radius within which the mean density is 200 times the mean density of the universe, $\bar{\rho}$. As we try to connect the mass accretion rate of dark matter haloes to star formation, we define halo masses as the total mass within $R_{200\mathrm{m}}$,
\begin{equation}\label{eq:M200}
    M_h = 200 \frac{4\pi}{3} R^3_{200\mathrm{m}} \bar{\rho},
\end{equation}
where $\bar{\rho}(t) = {\rho}_0 a(t) ^ {-3}$.

We assume that during gravitational collapse, the gas experiences strong shocks and thermalises its kinetic infall energy to the virial temperature of the halo,
\begin{equation}\label{eq:T_vir}
	T_\mathrm{vir} = \frac{\mu m_p G M_h}{5 k_\mathrm{B} R_{200\mathrm{m}}},
\end{equation}
where we have assumed a uniform cloud of monatomic gas. $M_h$ is the mass of the halo, $\mu$ is the mean molecular weight of the gas in the halo, which we have assumed $\mu \approx 0.6$ for a fully ionized plasma of primordial composition, $k_\mathrm{B}$ is the Boltzmann constant, and $m_p$ the proton mass. Note from \cref{eq:M200} that for a given halo mass, the radius of the halo, $R_{200\mathrm{m}}$, changes with time as the mean density of the Universe evolves. 

\begin{figure}
\centering 
\includegraphics[width=0.48\textwidth]{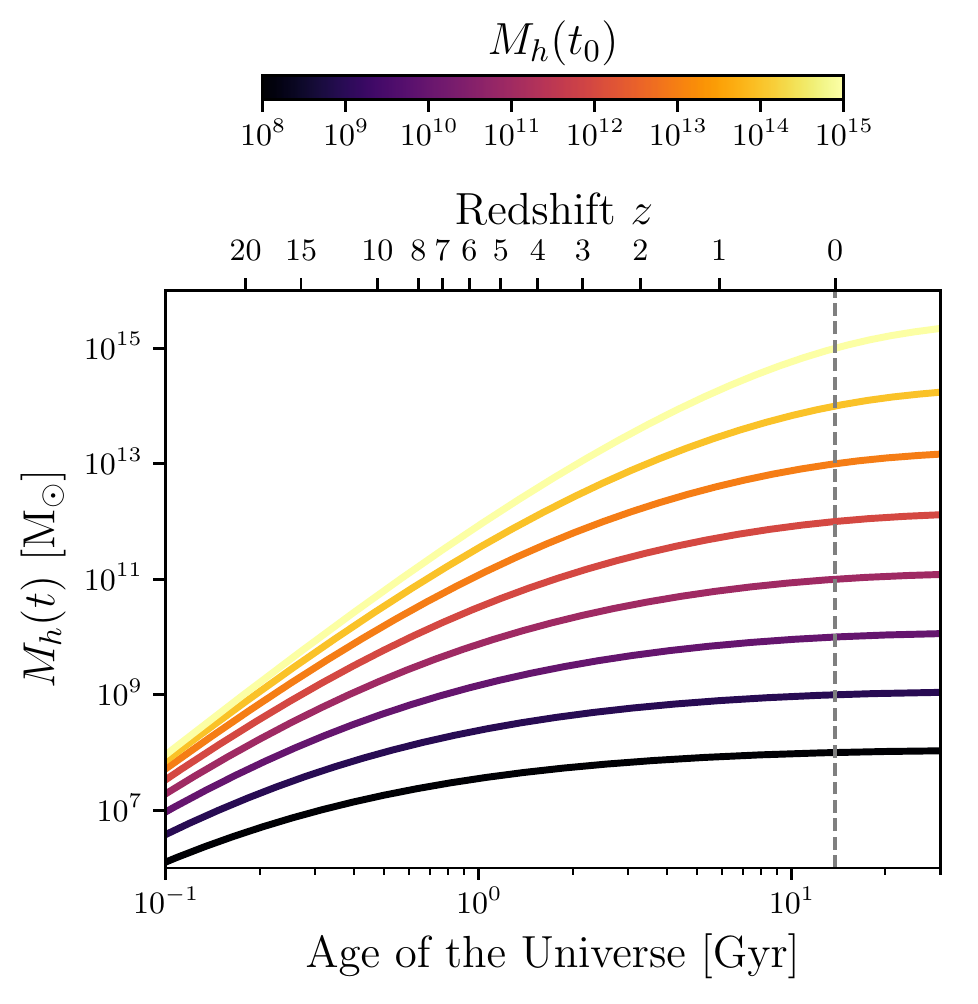}
 \vspace{-1.5em}
 \caption{Average halo mass as a function of cosmic time derived in \cref{eq:halo_mass}. A model for the cosmological parameters for a standard $\Lambda$CDM universe as inferred by the \citet{planck_collaboration_planck_2014} is shown with solid lines. Colour coding represents different halo masses, $M_0$, at the present cosmic time $t_0$, $M_0=M_h(t_0)$.}
 \label{fig:halo_mass}
\end{figure}

\begin{figure*}
\centering 
\includegraphics[width=0.95\textwidth]{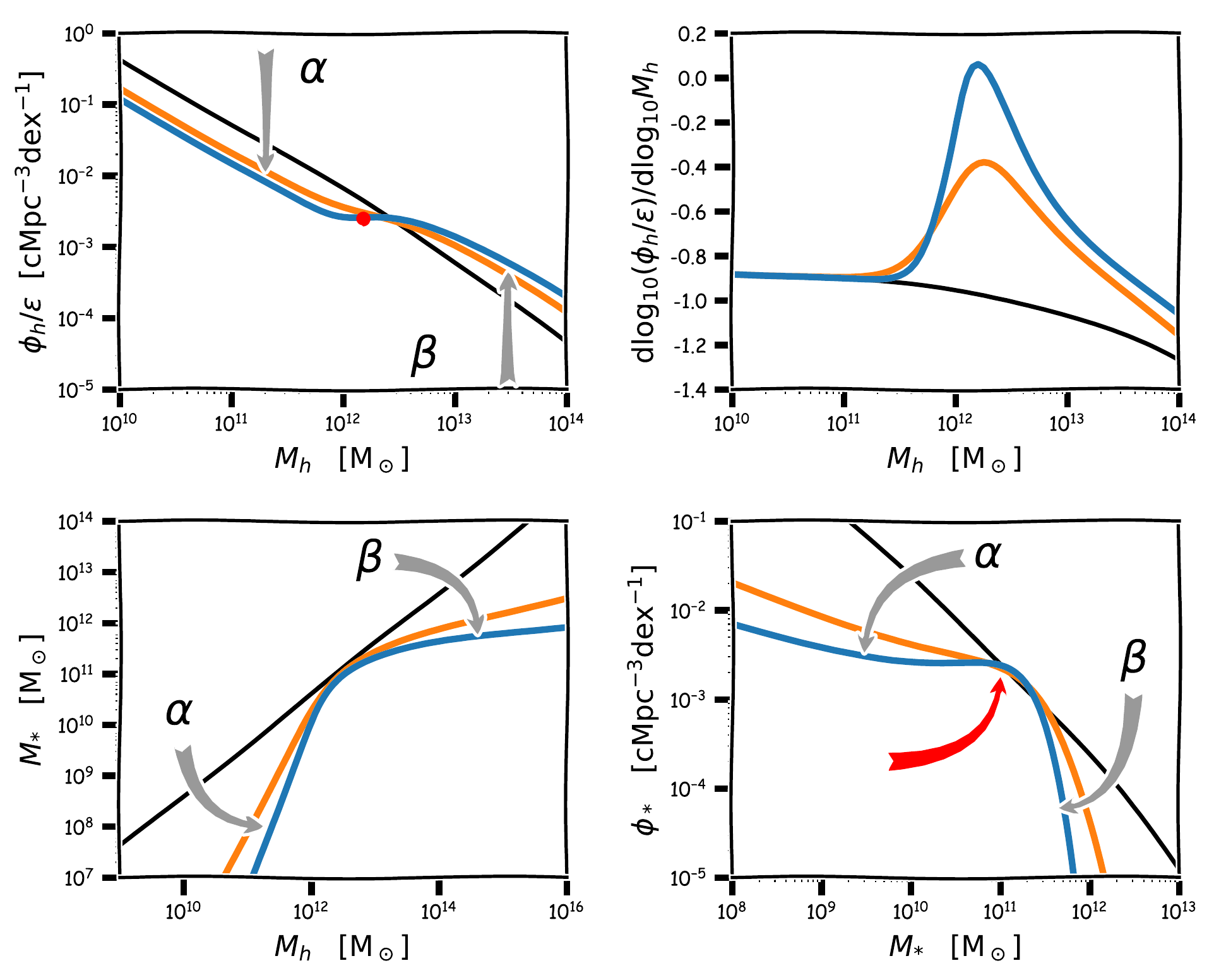}
  \vspace{-1.5em}
 \caption{A schematic illustration of the role played by the low-mass and high-mass end slopes of the SHMR in shaping the GSMF (see \cref{eq:GSMF}). Two arbitrary models are shown. A model with both $\alpha$ and $\beta$ large is shown in blue. The orange line illustrates a model with smaller $\alpha$ and $\beta$. \textit{Top-Left:} Product of the halo mass function and the inverse of the logarithmic slope of the SHMR, $\varepsilon$, given in \cref{eq:logslope}. The halo mass function is shown in black. For the blue line, the low-mass end is multiplied by a factor $1/(1+\alpha)$, while the high-mass end is multiplied by a factor of $1/(1-\beta)$. As both $\alpha$ and $\beta$ are positive, this creates an inflection point (shown as a red dot) in the distribution. \textit{Top-Right:} Logarithmic derivative of $(\phi_h / \varepsilon)$. The different changes in the normalisation cause a maximum in the distribution. Hence, there is an inflection point as the second derivative vanishes and changes sign at $\sim M_\mathrm{crit}$. The black line shows the logarithmic slope of the halo mass function. \textit{Bottom-Left:} The SMHR is shown. The black line shows a relationship of $M_* \propto M_{h}$. At low masses, SFR is suppressed because of the efficiency of feedback from star formation, yielding a slope of $(1+\alpha)$. At higher masses a slope of $(1-\beta)$ is expected as cooling of the inflowing gas is suppressed by heating from BHs. \textit{Bottom-Right:} The GSMF is shown. The black line shows a relationship of $\phi_* \propto \phi_{h}$. The low-mass and high-mass end slopes of the SHMR suppress the abundance of low and high mass galaxies respectively, but also create a characteristic ``bump'' at the knee of the GSMF.} 
 \label{fig:Origin_GSMF}
\end{figure*}

\subsection{The model}\label{sec:themodel}

The analytical galaxy formation model developed here is comprised of three main components, which are summarised in the schematic diagram of \cref{fig:schematic}: \begin{itemize}
    \item A cosmological model (blue block)
    \item An astrophysical model that sets the relation between galaxies and their haloes (green block)
    \item The galaxy formation outputs (orange block)
\end{itemize}

By using the Taylor expansion solution of the Friedmann equations introduced by \cite{salcido_impact_2018}, the formation and evolution of dark matter haloes 
can be described analytically. This component is shown as the blue block in \cref{fig:schematic}. The growth rates of haloes depend on the
cosmological parameters ${\rho}_0$, $\Lambda$, $H_0$ and the shape of the matter density fluctuation power-spectrum. We parametrise the variance of the spherically averaged smoothed density field, $S = \sigma^2$, as a power law $S\approx S_0 (M_h/10^{12} \mathrm{M}_\odot)^{-\gamma}$ with slope ${\gamma}$. Since we are interested in only a small range of halo mass, this is a sufficiently accurate description. The derivation of these equations are presented in \Cref{sec:derivation}. For convenience, we define the cosmology dependent approximations for the equations that appear bellow using the function,
\begin{equation}\label{eq:f_lambda}
    f_\Lambda(t, A, B)  = 1 + A \left(\frac{t}{t_\Lambda}\right)^2 + B \left(\frac{t}{t_\Lambda}\right)^4,
\end{equation}
where $t_\Lambda = \sqrt{{3}/{\Lambda c^2}}$, $\Lambda$ is the value of the cosmological constant, and the coefficients $A$ and $B$ are obtained by using the Taylor expansion solution of the Friedmann equations in \cite{salcido_impact_2018}.

Astrophysical processes (shown in green in \cref{fig:schematic}), enter into the model through the effective star formation efficiency, which is fully described by the efficiency $\epsilon_*$ (\cref{eq:eps_star}). We consider two models, in which $\epsilon_*$ is a function of halo mass or virial temperature. This component of the model is described by four parameters, $M_\mathrm{crit}$ (or $T_\mathrm{crit}$), $\epsilon_\mathrm{N}$, $\alpha$ and $\beta$ following \cref{eq:epsilon} or \cref{eq:epsilon_T_vir}. 

In order to simplify the numerical constants in the equations presented in this section, we have substituted the numerical values for the cosmological parameters for a standard $\Lambda$CDM universe as inferred by the \citet{planck_collaboration_planck_2014}, i.e. $\Omega_\mathrm{m} {=} 0.307$, $\Omega_\Lambda {=} 0.693$, $\Omega_b {=} 0.04825$, $H_0{=} 67.77 \, \mathrm{km} \,\mathrm{s}^{-1} \mathrm{Mpc}^{-1}$, $\rho_0 = 3.913\times 10^{10} \Msol \Mpc^{-3}$, $\gamma {=} 0.3$, $t_0 {=} 13.8 \Gyr$ and $S_0 {=} 3.98$. The full cosmology dependence of the numerical constants is given in \cref{sec:derivation}, and are highlighted using a coloured superscript $(^{\textcolor{red}{c^*}})$.

\begin{itemize}
	\item \textit{Halo mass history}
\end{itemize}
The analytic form of the growth rate equations allows us to simply describe the growth of haloes as a function of their present-day mass, $M_0$:
\small
\begin{equation}\label{eq:halo_mass}
    \begin{aligned}
& \frac{M_h(t)}{10^{12}\mathrm{M}_\odot}  =  \\
& \left\{\left(\frac{M_0}{10^{12}\mathrm{M}_\odot}\right)^{-\gamma /2} + 0.31 \gamma \, 
    \left[\left(\frac{t}{t_m}\right)^{-2/3}f_\Lambda(t, 0.16, - 0.01) - 1.67 \right]\right\}^{-2/\gamma}
    \end{aligned}
\end{equation}
\normalsize
where $t_\mathrm{m} = \sqrt{{3}/{8\pi G{\rho}_0}}$, and ${\rho}_{0}$ is the mean matter density of the Universe at the present time. For the \citet{planck_collaboration_planck_2014} cosmological parameters, $t_m {=} 26.04 \Gyr$ and $t_\Lambda {=} 17.33 \Gyr$. \Cref{fig:halo_mass} shows the individual mass histories for haloes of a given mass $M_0$ at the present cosmic time (represented by the colour coding). 

\begin{itemize}
    \item \textit{Halo mass function}
\end{itemize}

Using the Press \& Schechter formalism \citep{press_formation_1974}, the co-moving abundance of haloes of mass $M_h$ at time $t$ is given by,
\small
\begin{equation}\label{eq:halo_mass_fun}
    \begin{aligned}
         \frac{\dd n(M_h,t)}{\mathrm{dlog}_{10} M_h} & = 5.43\times 10^{-3}
        \,\mathrm{cMpc}^{-3} \, \left(\frac{M_h}{10^{12}\mathrm{M}_\odot}\right)^{-\left(1-\frac{\gamma}{2}\right)} \,\, \times \\
        &\left(\frac{t}{t_m}\right)^{-2/3} 
        f_\Lambda(t, 0.16, - 0.01) \,\, \times \\ 
         & \exp \left[ -0.13 \left(\frac{M_h}{10^{12}\mathrm{M}_\odot}\right)^{\gamma}
         \left(\frac{t}{t_m}\right)^{-4/3} f_\Lambda(t,0.32, 0) \right].
    \end{aligned}
\end{equation}
\normalsize
This equation consists of two parts, a low-mass power law dependence close to $M_h^{-1}$, and an exponential cut-off at high masses. For a given halo mass, the 
abundance initially increases as the exponential suppression is reduced, but at late times the halo abundance slowly decreases because of the power-law term.

\begin{itemize}
	\item \textit{The galaxy stellar mass function and the origin of the Schechter function}
\end{itemize}

The galaxy stellar mass function (GSMF) has been reasonably well measured over much of cosmic time, so that, for a know cosmology, the GSMF provides a good measurement
of the efficiency by which haloes convert their baryons into stars. Typically, the GSMF (\cref{eq:GSMF}) is parameterised by a Schechter function \citep[][]{schechter_analytic_1976},
\begin{equation}\label{eq:single_schechter}
    \begin{aligned}
        \phi(M) =  \phi^* \qty (\frac{M}{M^*})^{\alpha} e^{-{M}/{M^*}},
    \end{aligned}
\end{equation} 
where $\phi^*$ provides the normalisation, and $M^*$ is a characteristic galaxy stellar mass where the power-law form of the function cuts off. The form of this
function was originally motivated by the halo mass dependence given in \cref{eq:halo_mass_fun}. Importantly, however, the shape GSMF is only indirectly related to the halo mass function (eg., \citet{benson_what_2003}), with observations
showing that the power-law slope is much flatter than that expected for the
halo mass function. Moreover, recent measurements of the GSMF at low redshift \citep[e.g.][]{baldry_galaxy_2008,li_distribution_2009,baldry_galaxy_2012,moustakas_primus:_2013}, have proven that a single Schechter function is insufficient to describe the density of galaxies. Specifically, the low redshift GSMF shows a clear evidence for a low-mass upturn, or equivalently, a "pile-up" in the abundance of galaxies 
around $M^*$. Typically, a double Schechter function parametrisation has been used to better describe observational data,
\begin{equation}\label{eq:double_schechter}
    \begin{aligned}
        \phi(M) =  \qty [\phi_1^* \qty (\frac{M}{M^*})^{\alpha_1} + \phi_2^* \qty (\frac{M}{M^*})^{\alpha_2}] e^{-{M}/{M^*}}.
    \end{aligned}
\end{equation} 

In the model presented here, the GSMF can be computed as a function of time, by combining the halo mass function (\cref{eq:halo_mass_fun}) and the efficiency of star formation through \cref{eq:GSMF} and \cref{eq:logslope}. These equations link the observed shape of the GSMF, to the underlying dark matter halo distribution, and hence to the cosmological background. They also link galaxies to their dark matter haloes, providing valuable information about the efficiency by which haloes convert their baryons into stars.

Further consideration shows that they provide a description of the non-trivial shape of the GSMF and the need for a double Schechter function to describe it. While the underlying distribution of dark matter haloes is theoretically predicted to be a single Schechter function \citep{press_formation_1974}, its transformation to the GSMF relies on \cref{eq:GSMF}. When the halo mass function is multiplied by the inverse of the logarithmic slope of the SHMR, the low-mass end is multiplied by a factor $1/(1+\alpha)$, while the high-mass end is multiplied by a factor of $1/(1-\beta)$. As both $\alpha$ and $\beta$ are positive, this creates an a kink in the gradient, shown with a red dot in the top-left panel of \cref{fig:Origin_GSMF}. These different changes in the normalisation cause a maximum in the logarithmic derivative of $(\phi_h / \varepsilon)$ shown top-right panel of the figure. Hence, there is an inflection point in the distribution, as the second derivative vanishes and changes sign at $\sim M_\mathrm{crit}$. At this point, the abundance of galaxies decreases slowly, or even rises, as a function of mass, creating a ``bump'' at the knee of the GSMF. The sharper the transition (ie., the larger $\alpha+\beta$), the more pronounced the bump at the knee of the GSM. These effects are illustrated in \cref{fig:Origin_GSMF}. Physically, this can interpreted as galaxies of similar masses ``piling up'' at the peak star formation efficiency, i.e. $M_h \approx M_\mathrm{crit}$, as they rapidly stop forming many more stars. In \cref{sec:impact}, we will systematically vary the four parameter in the efficiency model in \cref{eq:epsilon} to investigate their effect on different galaxy formation outputs. 

\begin{itemize}
    \item \textit{The galaxy stellar mass growth}
\end{itemize}

We now have all the necessary ingredients to calculate the stellar mass growth of individual galaxies through cosmic time. Substituting \cref{A:eq:mdot_eps_series} into \cref{eq:eps_star}, the stellar mass is given by the integral of,
\small
\begin{equation}\label{eq:stellar_mass_int}
    \begin{aligned}
        \frac{\dd {M_*}}{\dd t} & =\epsilon_* f_b \, \left[ \frac{1}{M_h} \frac{\dd M_h}{\dd t} \right] M_h \\
        & =\epsilon_* f_b \, \, 1.6  \times 10^{10} \mathrm{M}_\odot \Gyr^{-1} \left(\frac{t}{t_m}\right)^{-5/3} \,\, \times \\
        & f_\Lambda(t, -0.32, 0.06) \left(\frac{M_h}{10^{12}\mathrm{M}_\odot}\right)^{\left(1+\frac{\gamma}{2}\right)}
        .
    \end{aligned}
\end{equation}
\normalsize
where $\epsilon_*$ is given by either \cref{eq:epsilon} or \cref{eq:epsilon_T_vir}.

Assuming an instantaneous recycling approximation \citep{schmidt_rate_1963}, the relation between the stellar mass growth due to star formation and the observed galaxy SFR is simply given by,
\begin{equation}\label{eq:mass_loss}
\begin{aligned}
    \mathrm{SFR} &= \frac{\dot{M}_*}{(1-R)},
\end{aligned}{}
\end{equation}
where $R$ is the fraction of mass of gas that is instantaneously returned into the interstellar medium by an entire stellar generation. For a universal \cite{chabrier_galactic_2003} initial mass function (IMF), $R=0.41$. 

Furthermore, both \textit{in situ} star formation and galaxy mergers contribute to the total stellar build up of galaxies. In low mass haloes, most of the stellar build up is expected to come from \textit{in situ} star formation, while the most massive galaxies experience almost no internal star formation and grow mainly by mergers with smaller satellite galaxies. Hence, the fractional contribution of accreted stars to the total stellar mass build up of galaxies is a steep function of halo mass \citep[e.g.][]{Rodriguez_Gomez_2016,qu_chronicle_2017,pillepich_first_results_2018}. Assuming that all of the stellar growth of haloes of mass $M_\mathrm{crit}$ and bellow is due to internal star formation, we parametrise the fraction of stellar mass growth from \textit{in situ} SFR by a broken power law as,
\begin{equation}\label{eq:in_situ}
\begin{aligned}
    f_\mathrm{SFR} =
    \begin{cases}
         1  &\text{   for } M_h \leq M_\mathrm{crit} \\
         \qty(\frac{{M_h}}{M_\mathrm{crit}})^{\eta} & \text{   for } M_h > M_\mathrm{crit},
    \end{cases}
\end{aligned}
\end{equation}
where $M_\mathrm{crit}$ is the effective star formation peak efficiency defined in \cref{eq:epsilon}. For the virial temperature-dependent efficiency model (Model II), $M_\mathrm{crit}$ also varies with time, and can be calculated using the critical virial temperature in \cref{eq:T_vir}. We fix the value of $\eta$ by assuming that at redshift $z=0$, where $M_\mathrm{crit}\approx 10^{12} \mathrm{M}_{\odot}$, $f_\mathrm{SFR}(10^{13} \mathrm{M}_{\odot})$ is $\sim 50\%$ \citep{pillepich_first_results_2018}, hence,
\begin{equation}
    \eta = \log_{10}(0.5) / (13 - 12) = - 0.3.
\end{equation}

Putting \cref{eq:mass_loss,eq:in_situ} together, the fraction of stellar mass growth of central galaxies due to \textit{in situ} formation is given by,
\begin{equation}\label{eq:SFR_obs}
\begin{aligned}
    \mathrm{SFR} &= \frac{\dot{M}_*}{(1-R)} \times f_\mathrm{SFR}. 
\end{aligned}{}
\end{equation}

\begin{itemize}
	\item \textit{The cosmic star formation rate density}
\end{itemize}
The total cosmic SFR density is given by the integral of all star formation in all haloes,
\small
\begin{equation}\label{eq:rho_int}
\begin{aligned}
    \dot{\rho}_{_\mathrm{SFR}}(t) &= \int \dot{M}_* \frac{f_\mathrm{SFR}}{(1-R)}  \frac{\dd n(M_h,t)}{\mathrm{dlog}_{10} M_h} \mathrm{dlog}_{10} M_h \\
    &= \int \epsilon_* f_b \dot{M}_h \frac{f_\mathrm{SFR}}{(1-R)}  \frac{\dd n(M_h,t)}{\mathrm{dlog}_{10} M_h} \mathrm{dlog}_{10} M_h.
\end{aligned}
\end{equation}
\normalsize
Using the stellar mass growth rate from \cref{eq:stellar_mass_int}, the halo mass function from \cref{eq:halo_mass_fun}, together with the effective star formation efficiency from \cref{eq:eps_star}, the contribution to the cosmic SFR density from haloes of mass $M_h$ (the integrand of \cref{eq:rho_int}) is given by,
\small
\begin{equation}\label{eq:ssfr_rate_density_dm}
    \begin{aligned}
        \frac{\dd \dot{\rho}_{_\mathrm{SFR}}}{\mathrm{dlog}_{10} M_h} & = \epsilon_* f_b \frac{f_\mathrm{SFR}}{(1-R)} \, 8.7 \times10^{7} \Msol \Gyr^{-1} \mathrm{cMpc}^{-3} \,\, \times \\
         &\left(\frac{M_h}{10^{12}\mathrm{M}_\odot}\right)^{\gamma} \left(\frac{t}{t_m}\right)^{-7/3} f_\Lambda(t,-0.16, 0) \,\,\times \\
         & \exp \left[-0.13 \left(\frac{M_h}{10^{12}\mathrm{M}_\odot}\right)^{\gamma}\left(\frac{t}{t_m}\right)^{-4/3} f_\Lambda(t, 0.32, 0) \right],
    \end{aligned}
\end{equation}
\normalsize
where $\epsilon_*$ is modelled using \cref{eq:epsilon} or \cref{eq:epsilon_T_vir}. The differential form of \cref{eq:ssfr_rate_density_dm} explicitly shows the contribution from haloes of different masses $M_h$, to the total cosmic SFR density. 

\Cref{eq:GSMF,eq:halo_mass,eq:halo_mass_fun,eq:stellar_mass_int,eq:SFR_obs,eq:ssfr_rate_density_dm}, together with a model of the effective star formation efficiency, \cref{eq:epsilon} or \cref{eq:epsilon_T_vir}, provide a full mathematical framework to explore the effects of cosmology and baryonic physics on galaxy formation. In the next section, we will explore the effect of the different efficiency parameters on the galaxy SFR, GSMF and the cosmic SFR density.

\begin{figure}
\centering 
\includegraphics[width=0.48\textwidth]{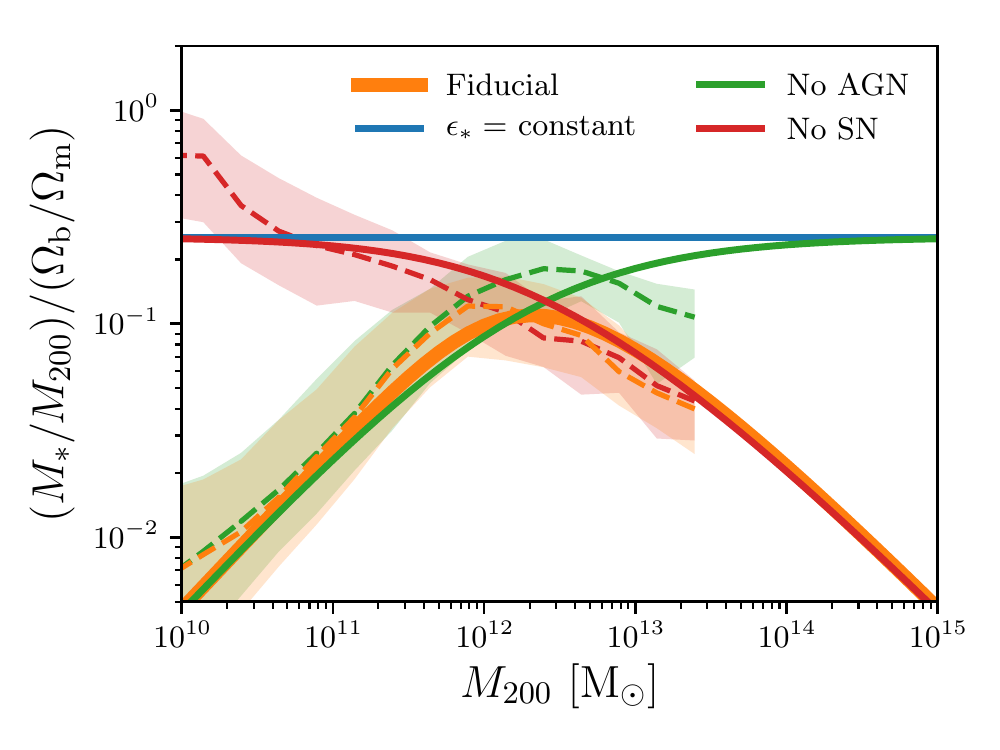}
  \vspace{-1.5em}
 \caption{Median stellar-halo mass ratio for central galaxies for three variations of the \textsc{eagle} $(50\mathrm{cMpc})^{3}$ simulations at redshift $z{=}0$ (dashed lines), compared to their equivalent analytic effective star formation efficiency model (solid lines). The orange line shows the Ref-L050N0752 \textsc{eagle} model \citep{schaye_eagle_2015,crain_eagle_2015}, which uses the same calibrated sub-grid parameters as the reference model $(100\mathrm{cMpc})^{3}$, ran with the same resolution, but in a smaller volume. The ``No AGN'' run (green) uses the same calibrated sub-grid parameters as the reference model but removing feedback from BHs. For the ``No SN'' model (red), feedback from star formation has been removed. We note that there is no \textsc{eagle} equivalent to the ``constant'' (or ``no feedback'') model. The faint shaded regions enclose the 10th–90th percentiles. While much more computationally expensive, the behaviour of the full hydrodynamical simulations is well approximated by the analytic models introduced here.}
 \label{fig:SHMR_Eagle}
\end{figure}

\begin{figure}
\centering 
\includegraphics[width=0.48\textwidth]{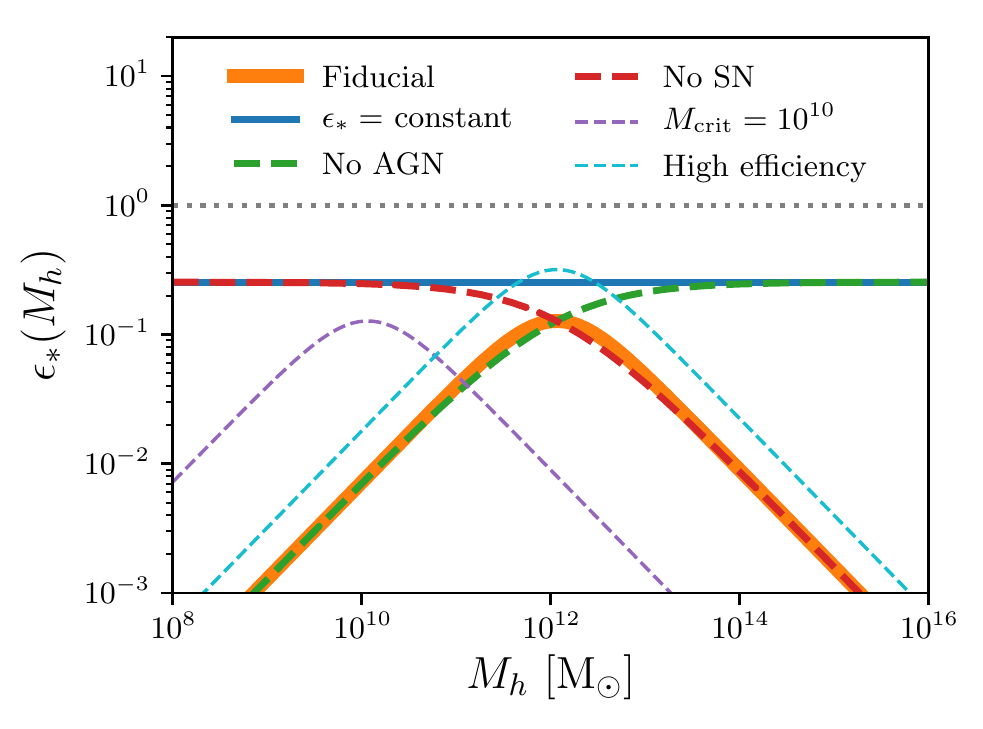}
 \vspace{-1.5em}
 \caption{The effective star formation efficiency $\epsilon_*$ as a function of halo mass for the six models described in \cref{tab:models}. For the ``Fiducial'' model, the efficiency peaks at masses similar to Milky-Way sized haloes ($10^{12} \Msol$) and fall steeply for higher and lower masses with $\alpha=0.75$ and $\beta=0.75$. For the ``Constant'' model, a fixed fraction of the baryon budget is turned into stars, regardless of the halo mass. The ``No AGN'' model describes a scenario where the efficiency of feedback process is weak for massive objects. The ``No SN'' model describes a scenario where the efficiency of feedback process is weak in small haloes. The ``$M_\mathrm{crit} = 10^{10}$'' model explore the effect of changing the critical halo mass. The ``High efficiency'' model has the same slopes as the fiducial model, but with a higher normalisation. A 100\% efficiency is shown with a grey dotted line.}
 \label{fig:epsilon_models}
\end{figure}

\section{The impact of the effective star formation efficiency}\label{sec:impact}

We now use our model to explore the effect of the different efficiency parameters in the galaxy formation outputs in the orange block of \cref{fig:schematic}. It is common to characterise galaxy properties over halo masses, and for simplicity, in this section we will only use a halo mass-dependent star formation efficiency model (i.e. $\epsilon_*$ is constant across cosmic time). 

It has been estimated that the SHMR peaks at masses similar to Milky-Way sized haloes (${\sim} 10^{12} \Msol$). Typically, at low masses, the SFR is suppressed because of the efficiency of stellar feedback. On the other hand, at higher masses the cooling of the inflowing gas is suppressed by heating from supermassive BHs \citep[e.g.][]{white_galaxy_1991, bower_breaking_2006,benson_g_2012}. The ``Fiducial'' model captures this behaviour with both $\alpha$ and $\beta$ being positive and equal to 0.75. 

\begin{table}
\centering
\caption{Effective star formation efficiency parameters for the six idealised models. To agree with observational data, the values of $\alpha$ and $\beta$ are typically positive. The ``Fiducial'' model captures this behaviour while the five variations systematically explore the effect of the effective star formation efficiency on the physics of galaxy formation.}
\label{tab:models}
\begin{tabular}{lcccc}\hline
& \multicolumn{1}{c}{\textbf{$\epsilon_\mathrm{N}$}} & \multicolumn{1}{c}{\textbf{$M_\mathrm{crit}$}} & \multicolumn{1}{c}{\textbf{$\alpha$}} & \multicolumn{1}{c}{\textbf{$\beta$}} \\ \hline
\textbf{Fiducial} & 0.125 & $10^{12}$ & 0.75 & 0.75 \\  
\textbf{Constant} & 0.250 & $10^{12}$ & 0 & 0 \\  
\textbf{No AGN} & 0.125 & $10^{12}$ & 0.75 & 0 \\  
\textbf{No SN} & 0.125 & $10^{12}$ & 0 & 0.75 \\ 
\textbf{$M_\mathrm{crit} = 10^{10}$} & 0.125 & $10^{10}$ & 0.75 & 0.75 \\
\textbf{High Efficiency} & 0.320 & $10^{12}$ & 0.75 & 0.75 \\ \hline
\end{tabular}
\end{table}

We consider five alternative models varying the efficiency parameters systematically to explore the physics of galaxy formation. An extreme idealised case label as ``Constant'', describes a model where a fixed fraction of the baryon budget is turned into stars, regardless of the halo mass. The ``No AGN'' model describes a scenario where the efficiency of feedback processes is weak for massive objects. Physically, this could be thought as a model where feedback from active galactic nuclei is inefficient. The ``No SN'' model describes a scenario where the efficiency of feedback processes is weak in small haloes. Physically, this could be thought as a model where feedback from supernovae is inefficient. While much more computationally expensive, a similar behaviour to these models is reproduced in full hydrodynamical simulations (see \cref{sec:comparison} for a couple of examples). In \cref{fig:SHMR_Eagle} we show the median stellar-halo mass ratio for three variations of the \textsc{eagle} simulations where the subgrid prescription for stellar and AGN feedback have been removed. Indeed, our models capture the overall behaviour attained in simulations. An additional model labelled ``$M_\mathrm{crit} = 10^{10}$'' explores the effect of changing the critical, or transition, halo mass. A final model labelled ``High efficiency'' has the same slopes as the fiducial model, but with a different normalisation. We show in \cref{fig:epsilon_models} the effective star formation efficiency for the six models, and their parameters are summarised in \cref{tab:models}.

\begin{figure}
\centering 
\includegraphics[width=0.48\textwidth]{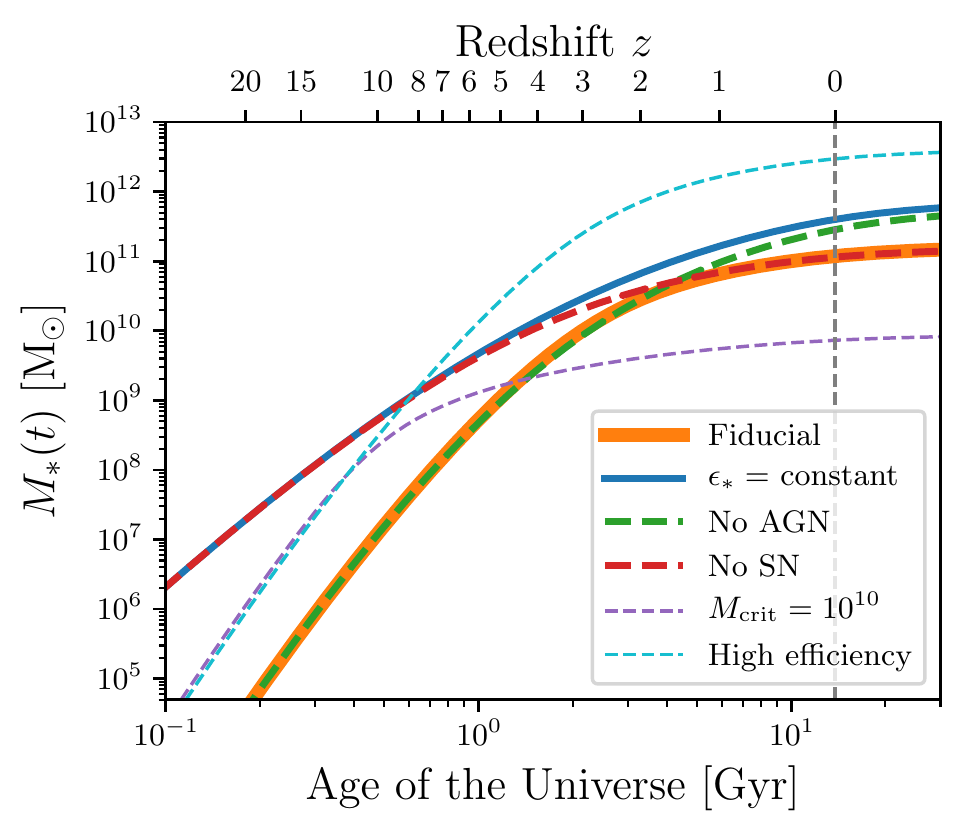}
 \vspace{-1.5em}
 \caption{An example of the evolution of the stellar mass in a halo of mass $M_0=10^{13}\mathrm{M}_\odot$ at the present time calculated by integrating \cref{eq:stellar_mass_int}. The different colours represent the different efficiency models. For the constant efficiency model, the stellar mass grows steadily with time tracking the mass assembly of the dark matter halo. For the fiducial, $M_\mathrm{crit} = 10^{10}\mathrm{M}_\odot$ and high efficiency models, the build up of stellar mass is faster (steeper slope), but once the corresponding critical halo mass is reached, the stellar mass plateaus and the halo hardly produces any additional stellar mass. The high efficiency model has the same shape as the fiducial model, but  a higher normalisation. As expected, the No SN and No AGN models build up more stellar mass before and after the halo reaches critical halo mass respectively.}
 \label{fig:m_star}
\end{figure}

\subsection{The build up of stellar mass}

First, we explore the effect of the star formation efficiency in the build up of stellar mass in individual haloes for the six models, which can be calculated by integrating \cref{eq:stellar_mass_int}.

 \cref{fig:m_star} shows an example of the evolution of the stellar mass in a halo of mass $M_0=10^{13}\mathrm{M}_\odot$ at the present time for the six efficiency models. For the constant efficiency model, the stellar mass grows steadily with time, and starts to slow down only at late times due to cosmic expansion, tracking the mass assembly of the dark matter halo. For the fiducial model, the build up of stellar mass is faster (steeper slope). Once the critical halo mass is reached ($M_\mathrm{crit} = 10^{12}\mathrm{M}_\odot$, corresponding to $M_* \approx 10^{10}\mathrm{M}_\odot$ for this model), the stellar mass plateaus. The No AGN model has a similar behaviour at early times, but once the critical halo mass is reached, star formation does not slow down and the halo reaches a higher stellar mass at the present time. On the other hand, the No SN model produces much more stellar mass at early times, but once the critical halo mass is reached, star formation slows down, and the halo reaches a similar final stellar mass as the fiducial model. The $M_\mathrm{crit}=10^{10}\mathrm{M}_\odot$ model presents a similar behaviour to the fiducial model, i.e. once the halo reaches the critical mass it hardly produces any additional stellar mass. However, as the critical mass is lower for this model, the transition happens at earlier times. The high efficiency model has the same shape as the fiducial model, but as expected, a higher normalisation.

\begin{figure}
\centering 
\includegraphics[width=0.48\textwidth]{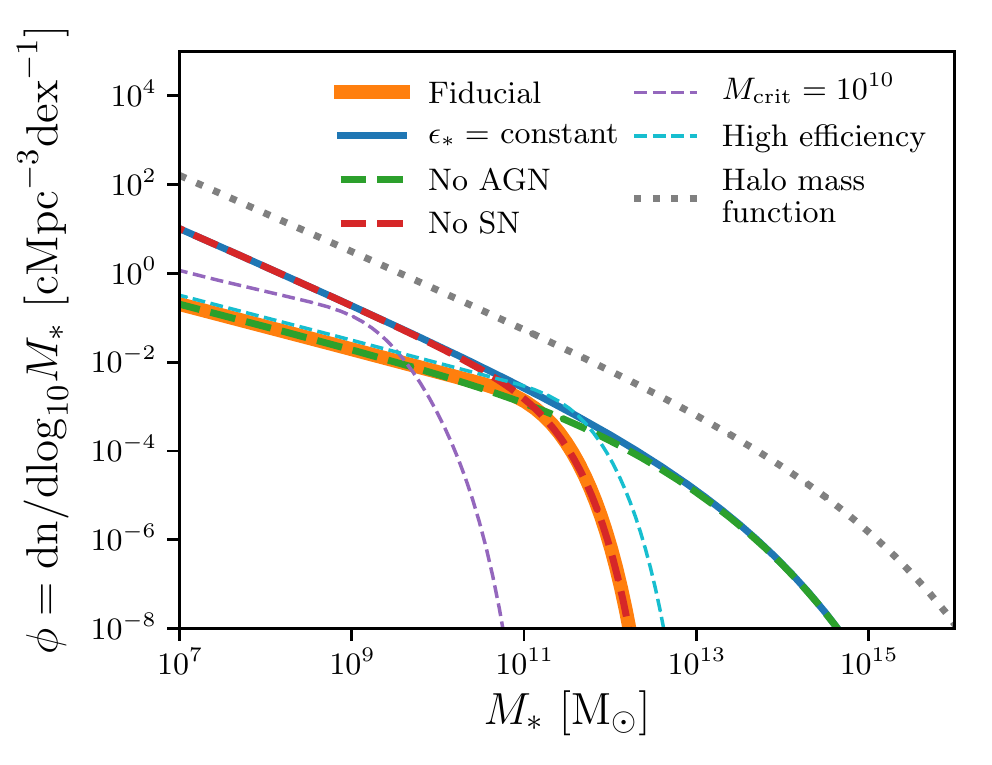}
  \vspace{-1.5em}
 \caption{The GSMF at the present time for the six efficiency models described in \cref{tab:models}. All models with $\beta=0.75$, i.e. at high masses cooling is suppressed by AGN feedback, exhibit a sharp cut-off at the critical halo mass (Fiducial, No SN, $M_\mathrm{crit} = 10^{10}\mathrm{M}_\odot$ and high efficiency). This shows that AGN feedback is mainly responsible for the characteristic knee of the GSMF. The location of the knee is determined both by the critical halo mass in the star formation efficiency, and the normalisation $\epsilon_\mathrm{N}$, as this causes a horizontal shift of the whole distribution. For the No SN model, the slope of the faint end of the GSMF is much steeper. If feedback process are inefficient both at the low mass and high mass end (a constant fraction of the baryon budget is turned into stars), the GSMF is identical to the halo mass function (dotted grey line) but shifted in mass by a constant value.}
 \label{fig:SMF}
\end{figure}

\subsection{The stellar mass function}

We use \cref{eq:GSMF,eq:mstar,eq:halo_mass_fun} to calculate the GSMF at redshift $z{=}0$ for the six efficiency models. We note that for a time evolving efficiency, \cref{eq:stellar_mass_int} should be used to calculate the stellar mass of any halo as a function of time. These equations allow us to obtain the SHMR and convolve it with the halo mass function to calculate the GSMF.  

\cref{fig:SMF} shows the GSMF at the present time for the six efficiency models. For the constant model, as it has been pointed out before \citep[e.g][]{benson_what_2003}, if feedback process are inefficient both at the low mass and high mass end, i.e. a constant fraction of the baryon budget is turned into stars in every halo, the GSMF does not exhibit the characteristic knee obtained in observations and is identical to the halo mass function (dotted grey line) but shifted in mass by a constant value. Once feedback processes are implemented, the location of the knee of the GSMF is determined by the critical mass in the star formation efficiency (fiducial, $M_\mathrm{crit} = 10^{10}$). Changing the critical mass also changes the normalisation of the distribution. All models with $\beta=0.75$, i.e. at high masses cooling is suppressed by AGN feedback, exhibit a sharp cut-off at the transition mass. Hence, AGN feedback is mainly responsible for the knee of the GSMF. The No AGN model has the same shallow slope at the faint end of the GSM function as the fiducial model, with a slight bend at high masses driven only by the exponential cut-off of the halo mass function. The No SN model presents the same knee as the fiducial model, but the slope of the faint end of the GSM function is much steeper. As discussed in \cref{sec:model}, the low-mass and high-mass end slopes of the SHMR produce a ``bump'' at the knee of the GSMF. Finally, the high efficiency model, as $\alpha$ and $\beta$ are the same as for the fiducial model, the shape of the GSMF is the same. i.e. the relative abundance of galaxies to their haloes, and hence the shape, is independent of the normalisation (as $\varepsilon$ does not depend on $\epsilon_\mathrm{N}$ in \cref{eq:logslope}). For a given halo mass, changing the normalisation maps that halo mass to a different galaxy mass. Hence, a change in the normalisation, $\epsilon_\mathrm{N}$, shifts the whole distribution only horizontally. In this case, the high efficiency model shifts the GSMF to the right compared to the fiducial model. 

\subsection{The cosmic SFR density}

The cosmic history of star formation is perhaps one of the most fundamental observables of our Universe. It has been observed to peak approximately 3.5 Gyr after the Big Bang ($z \approx 2$), and decline exponentially thereafter \citep[for a review see][]{madau_cosmic_2014}. Different groups have tried to model the complex physics driving the cosmic SFR by using, for example, full hydrodynamical simulations \citep[e.g.][]{schaye_eagle_2015,dave_mufasa:_2016,dubois_horizon-agn_2016,pillepich_simulating_2018}. Our analytic model disentangles the role of cosmology from the role of astrophysics, which in turn, allows us to examine the effect of the different efficiency parameters on the cosmic SFR density. 

We begin by noting that the behaviour of \cref{eq:ssfr_rate_density_dm} is governed by two main factors. First, a multiplier term that originates from both, the halo accretion rate, and the halo mass function, and is $\propto t^{-7/3}$. This, comes from the dynamical timescale of the universe getting larger. Second, an exponential term contribution due to the build up of haloes in the halo mass functions that is $\propto e^{-t^{-4/3}}$. For a given halo mass then, the exponential term dominates at early times, and the contribution to the cosmic SFR density is driven by the exponential build up of haloes. At late times, the exponential term asymptotically tends to a constant value, and the further evolution of the cosmic SFR is dominated by the multiplier term, i.e., it behaves as a power law. As discussed in \cite{salcido_impact_2018}, the contribution of dark energy is only relevant at late times, and at its observed value, it has a negligible impact on the star formation in the Universe.

\begin{figure}
\centering 
\includegraphics[width=0.48\textwidth]{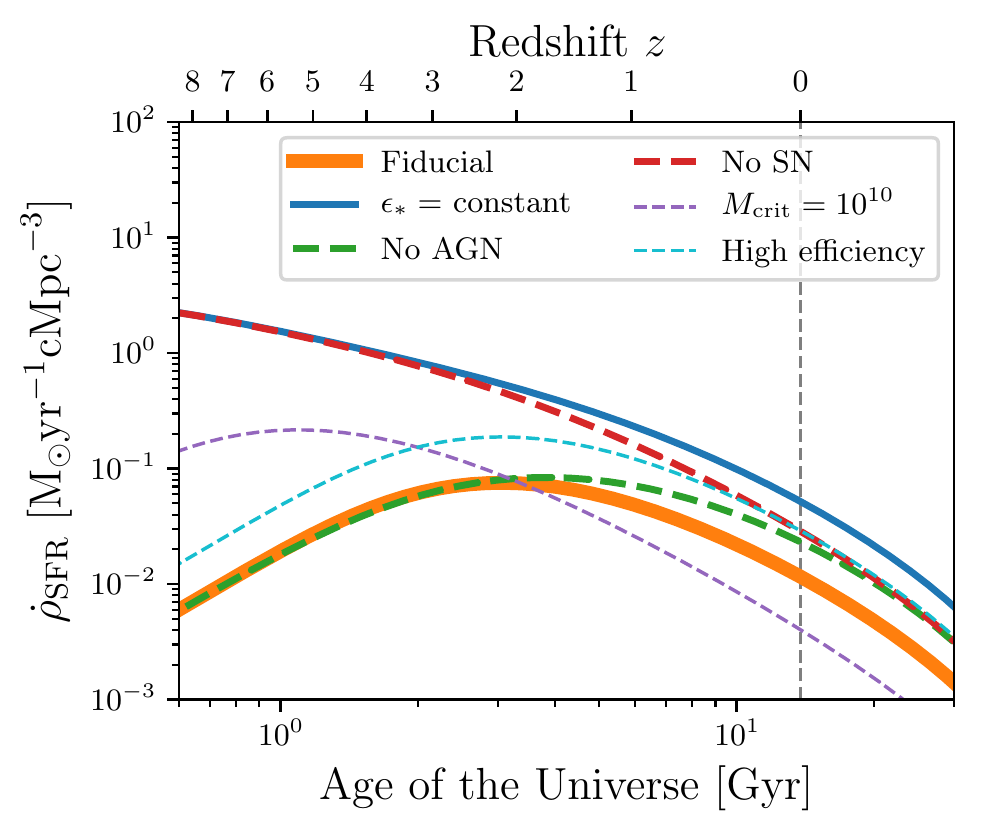}
  \vspace{-1.5em}
 \caption{The cosmic SFR density for the six efficiency models. As small haloes only contribute significantly at very early times, when the abundance of larger objects is strongly suppressed by the exponential term in the mass function, all models with $\alpha=0.75$, i.e. at low masses the SFR is suppressed because of the efficiency of stellar feedback, exhibit the characteristic peak in the observed cosmic history of star formation (Fiducial, No AGN, $M_\mathrm{crit} = 10^{10}\mathrm{M}_\odot$ and high efficiency). On the other hand, both models  with $\alpha=0$ (constant and No SN), do not exhibit the peak. This shows that supernovae feedback is mainly responsible for shaping the cosmic SFR density of the Universe. The figure shows that changing the transition mass $M_\mathrm{crit}$ has a great impact on the localisation and normalisation of the SFR peak, i.e. the SFR density is dominated by the largest haloes in which star formation is able to proceed without generating efficient feedback. AGN feedback only has a moderate effect on shaping the cosmic star formation, changing mildly its amplitude and localisation (No AGN model). The high efficiency model the same shape as the fiducial model, but with a higher normalisation.}
 \label{fig:CSFRD}
\end{figure}

\Cref{fig:CSFRD} shows the integrated cosmic SFR density for the six efficiency models computed using \cref{eq:ssfr_rate_density_dm}. For the fiducial model, while smaller haloes are more abundant than large objects, a smaller fraction of the inflowing material is converted into stars. As a result, the SFR density is dominated by the largest haloes in which star formation is able to proceed without generating efficient feedback. The smaller haloes only contribute significantly at very early times, when the abundance of larger objects is strongly suppressed by the exponential term in the mass function. We see therefore that the contribution of haloes of mass $\approx M_\mathrm{crit} = 10^{12}\Msol$, is representative of most of the SFR in the model. 

If star formation is efficient at all halo masses (constant model), then the cosmic SFR behaves like a power law with time, which only deviates from this behaviour at late times due to the suppression due to the cosmological constant. 

Examining the No SN model reveals the origin of the peak in the cosmic history of star formation is the efficient feedback in low mass galaxies. Without a mechanism to suppress star formation in small haloes, the history of the cosmic SFR density would not have its characteristic peak. Supernovae feedback is then mainly responsible for shaping the cosmic SFR density of the Universe. On the other hand, examining the No AGN model reveals that efficient feedback in high mass haloes only has a moderate effect on shaping the cosmic star formation. Without a mechanism to prevent star formation in massive galaxies, the cosmic SFR density would still exhibit a peak, only changing mildly its amplitude and localisation. However, the slope of the decline would be similar (orange vs green dashed lines).

Changing the transition mass $M_\mathrm{crit}$ has a great impact on the localisation of the SFR peak. As in the fiducial model, the contribution of haloes of mass $\approx M_\mathrm{crit}$, is representative of most of the SFR in the $M_\mathrm{crit} = 10^{10}$. Hence, the peak happens at earlier times, but also has a higher normalisation, as $10^{10} \mathrm{M}_\odot$ haloes are more abundant than $10^{12} \mathrm{M}_\odot$ haloes.

Finally for the high efficiency model, the shape of the SFR of the Universe is identical to the fiducial model, but with a higher normalisation. 

\section{Fitting observations}\label{sec:fit}

\begin{figure}
\centering 
\includegraphics[width=0.48\textwidth]{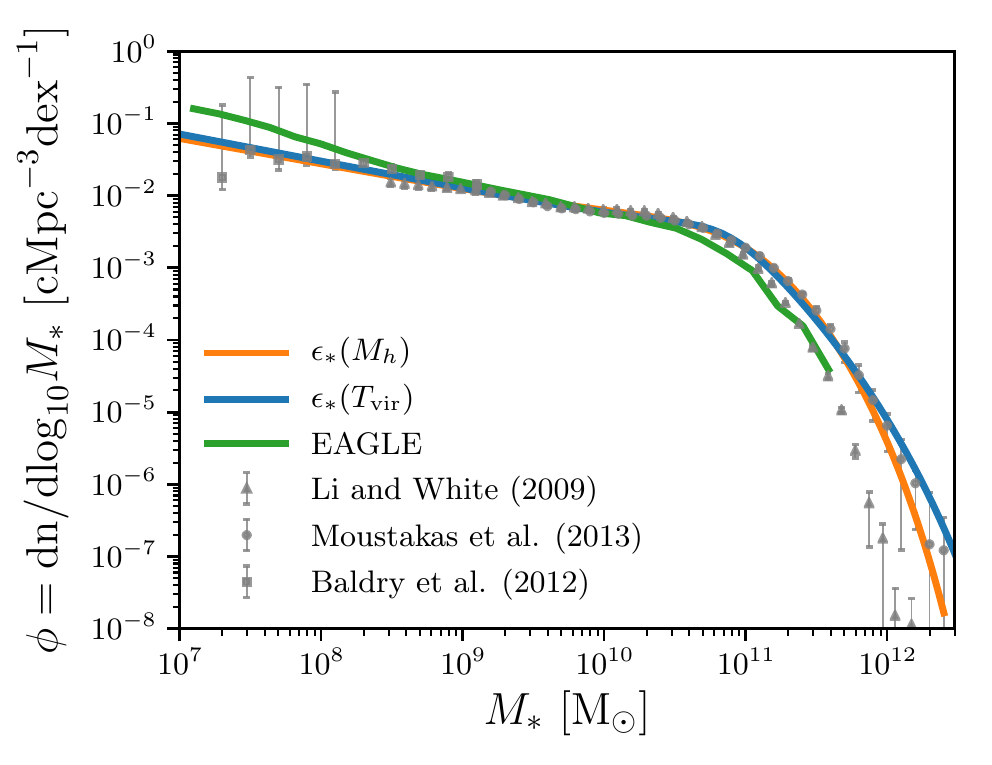}
  \vspace{-1.5em}
 \caption{Redshift $z=0.1$ GSMF for the best fit parameters for the halo mass-dependent model ($\epsilon_*({M_h})$) and the virial temperature-dependent model ($\epsilon_*({T_\mathrm{vir}})$). Observational data with their associated uncertainties from \protect\cite{li_distribution_2009, baldry_galaxy_2012,moustakas_primus:_2013} are shown with symbols. Both efficiency models provide a good fit to the present-day GSMF (see also the reduced $\chi^{2}$ statistics in \cref{tab:eps_mh}). Results from the \textsc{eagle} reference simulation are shown in green for reference.}
 \label{fig:GSMF_fit}
\end{figure}

In this section we compare the galaxy formation outputs from our analytic model with different observational datasets. We begin by calibrating our model to reproduce the GSMF at ${z}\sim0$ using observations from the Galaxy And Mass Assembly (GAMA) survey \citep{baldry_galaxy_2012} and the Sloan Digital Sky Survey (SDSS) \citep{moustakas_primus:_2013}\footnote{In this paper we used the standardised GSMF data from \cite{behroozi_universemachine:_2018}, which assumes a \cite{chabrier_galactic_2003} IMF, a \cite{bruzual_stellar_2003} stellar population synthesis model, dust corrections from \cite{calzetti_dust_2000}, and UV-stellar mass corrections.}. We use the reduced chi-squared statistic to derive the best-fitting effective star formation efficiency $\epsilon_*({M_h})$. Because the model is fully analytic, this calibration process is fast and easy to perform. \Cref{fig:GSMF_fit} shows the best fit model in orange. Results from the \textsc{eagle} reference simulation are shown in green for reference. The figure shows that a constant halo mass-dependent efficiency model provides an excellent fit to the present-day GSMF (with reduced $\chi_\nu^2$ = 1.5). The best best fit efficiency parameters are shown in \cref{tab:eps_mh}.

\begin{table}
\centering
\caption{Best fit parameters for the halo mass-dependent (\textit{Model I}, $\epsilon_*({M_h})$), and  virial temperature-dependent (\textit{Model II}, $\epsilon_*({T_\mathrm{vir}})$)  star formation efficiency models. For the virial temperature-dependent model, $M_\mathrm{crit}$ is given at redshift $z=0$, which corresponds to a critical virial temperature $T_{\mathrm{crit}} = 10^{5.3}\mathrm{K}$. As $T_{\mathrm{crit}}$ is kept constant, $M_\mathrm{crit} \propto a(t)^{3/2}$ (see \cref{sec:model1}). $\chi_\nu^2$ is the reduced chi-squared statistic used for goodness of fit testing.}\label{tab:eps_mh}
\begin{tabular}{lccccc}
\hline
\textbf{Model} & \textbf{$\epsilon_\mathrm{N}$} & \textbf{$M_\mathrm{crit}\,[\mathrm{M}_\odot]$}     & \textbf{$\alpha$}      & \textbf{$\beta$}      & \textbf{$\chi_\nu^2$} \\ \hline  
Mass-dependent & \multicolumn{1}{c}{\rule{0pt}{3ex}0.178}     & \multicolumn{1}{c}{$10^{11.68}$} & \multicolumn{1}{c}{1.537} & \multicolumn{1}{c}{0.656} & \multicolumn{1}{c}{1.5} \\ 
Temp-dependent & \multicolumn{1}{c}{\rule{0pt}{3ex}0.140}     & \multicolumn{1}{c}{$10^{12}$} & \multicolumn{1}{c}{2.377} & \multicolumn{1}{c}{0.834} & \multicolumn{1}{c}{1.6} \\ \hline
\end{tabular}
\end{table}

\begin{figure*}
\centering 
\includegraphics[width=0.95\textwidth]{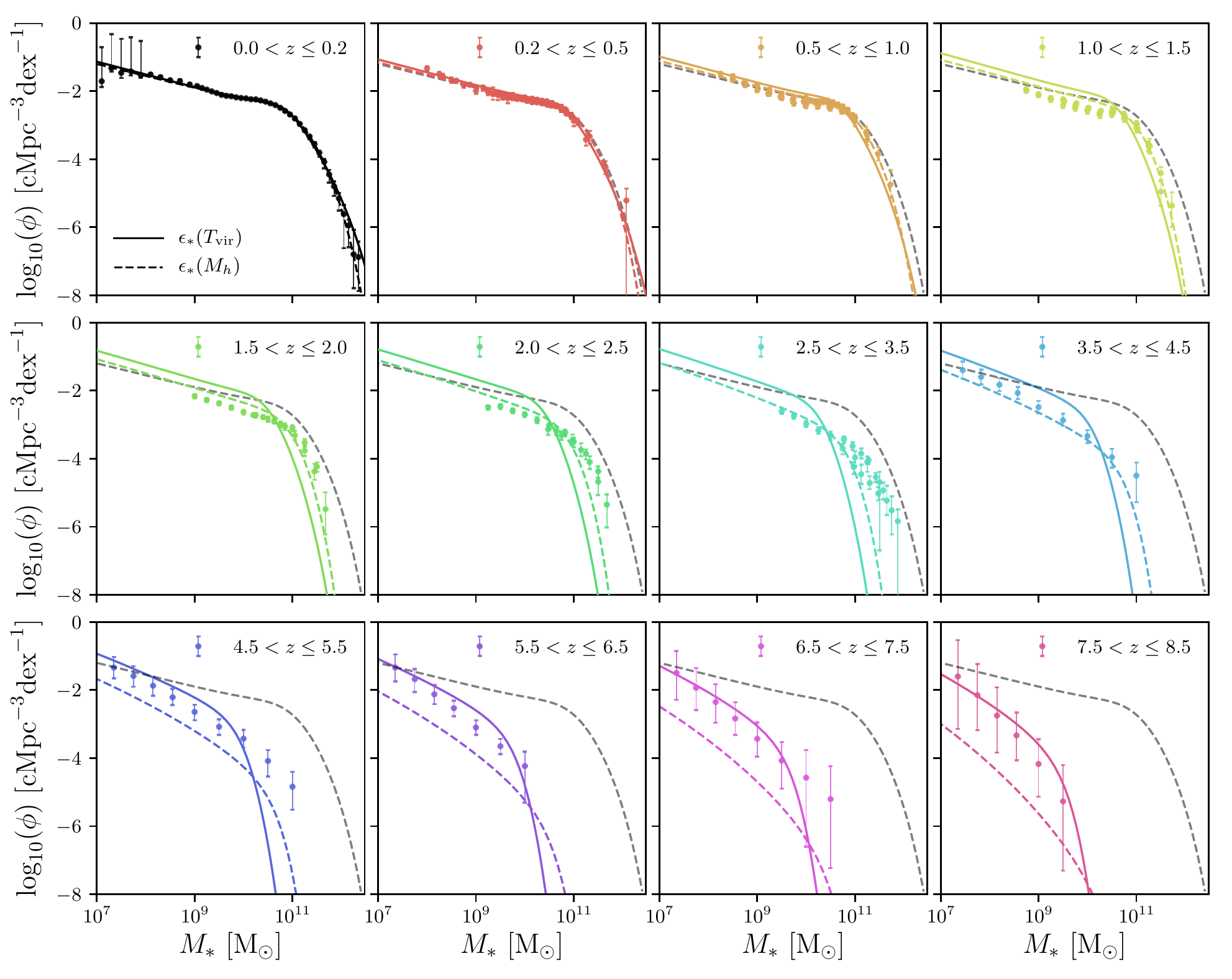}
  \vspace{-1.5em}
 \caption{Evolution of the predicted GSMF for the halo mass-dependent, and the virial temperature-dependent star formation efficiency models. Different panels and colours represent different redshifts. Observational data with their associated uncertainties from \protect\cite{baldry_galaxy_2012,moustakas_primus:_2013, tomczak_galaxy_2014, ilbert_mass_2013, muzzin_evolution_2013, song_evolution_2016} are shown by coloured symbols. The halo mass-dependent model is shown in dashed lines ($\epsilon_*({M_h})$). The virial temperature-dependent model is shown in solid lines ($\epsilon_*({T_\mathrm{vir}})$). The redshift $z \sim 0$ halo mass-dependent model is reproduced in each panel as a grey dashed curve, to highlight the evolution. While both models have been calibrated to reproduce the GSMF at $z \sim 0$, the halo mass-dependent model reproduces very well the evolution of the GSMF up to redshift $z=4$, but significantly under predicts the abundance of galaxies at higher redshift. On the other hand, the virial temperature model provides a good fit both at low and high redshift, but the evolution is too rapid at intermediate redshift ($z{=}1$ to $z{=}4$).}
 \label{fig:GSMF_evol}
\end{figure*}

\subsection{Contrasting halo mass and virial temperature efficiency models}\label{sec:t_vir}

Having established the best-fit efficiency parameters for the model, we can study the evolution of the model outputs. By construction, $\epsilon_*({M_h})$ is only a function of halo mass and is fixed in time. Hence, the evolution of the GSFM depends only on the evolution of the abundance of haloes of mass $M_h$ as a function of time, as described by the halo mass function. \Cref{fig:GSMF_evol} shows the evolution of the predicted GSMF for the halo mass-dependent star formation efficiency model in dashed lines. Different panels and colours represent different redshifts. Observational data from \cite{baldry_galaxy_2012,moustakas_primus:_2013,  tomczak_galaxy_2014, ilbert_mass_2013, muzzin_evolution_2013, song_evolution_2016}\footnote{In this paper we used the standardised GSMF data from \cite{behroozi_universemachine:_2018}, which assumes a \cite{chabrier_galactic_2003} IMF, a \cite{bruzual_stellar_2003} stellar population synthesis model, dust corrections from \cite{calzetti_dust_2000}, and UV-stellar mass corrections.} are shown with coloured symbols. 

Remarkably, a simple halo mass-dependent efficiency model reproduces very well the evolution of the GSMF up to redshift $z\approx4$. While the observed data at higher redshifts is highly uncertain, the halo mass-dependent model significantly under predicts the abundance of distant galaxies. This may hint to the need of a time-evolving efficiency model. As discussed in \cref{sec:model1},  it is reasonable to assume that the efficiency of star formation should be a function of the halo's virial temperature, which naturally evolves with cosmic time. In. this section, we investigate an effective star formation efficiency model that depends on virial temperate. 

\subsubsection{The three stages of galaxy formation}

An effective star formation efficiency model characterised by a time-independent critical virial temperature $T_\mathrm{crit}$ assumes that there exists a critical halo virial temperature at which there is a transition from where star formation driven outflows can escape, to where outflows stall inside the halo. Using the viral temperature of the halo to parameterise this tipping point, provides a natural evolution of the star formation efficiency. For a fixed halo mass, early collapsed haloes are more compact (denser), and one might expect a higher efficiency (for haloes with $T_\mathrm{vir} < T_\mathrm{crit}$).

In this simple picture, we can distinguish three stages of galaxy formation\footnote{While perhaps closely related, we distinguish these three \textit{stages} of galaxy formation from the three \textit{phases} in \cite{clauwens_three_2018}, as the latter refer mainly to a morphological evolution, rather than the entropy state and buoyancy of the gas.}, characterised by the virial temperature of the halo:

\begin{itemize}
    \item \textbf{Stellar feedback regulated stage:} star formation driven outflows effectively regulate the gas content of galaxies residing in haloes with virial temperature $T_\mathrm{vir} < T_\mathrm{crit}$. In this stage, efficient outflows prevent the density of central star forming gas building up, suppressing the growth of the central BH.
    \item \textbf{Efficient star forming/rapid growing black hole stage:} as haloes grow, the virial temperature increases to the point that the stellar outflows are no longer buoyant relative to their surroundings, and therefore stall (i.e  $T_\mathrm{vir} \approx T_\mathrm{crit}$). The density of gas builds up within the halo triggering high star formation rates and rapid BH growth. 
    \item \textbf{Black hole feedback regulated stage:} In haloes with $T_\mathrm{vir} > T_\mathrm{crit}$, the central BH is massive enough to produce efficient AGN feedback, in turn, regulating the gas content of the halo and preventing further star formation. 
\end{itemize}

An additional advantage of using the virial temperature to characterise the star formation efficiency, is that we can add a proxy for the effect of cosmic reionisation. Ultraviolet radiation from the first stars formed reionised neutral hydrogen, raising its entropy to a temperature of $\approx 10^4 \mathrm{K}$. This process prevented further cooling, hence preventing star formation in haloes with $T_\mathrm{vir} < 10^4 \mathrm{K}$ \citep{doroshkevich_origin_1967,couchman_pregalactic_1986,rees_lyman_1986,efstathiou_suppressing_1992,loeb_reionization_2001}. As a result of this suppression of star formation, only a fraction of the haloes with present-day mass $\approx 10^{10} \mathrm{M}_\odot$ form a galaxy, and no galaxies form below a halo mass of $\approx 10^7 \mathrm{M}_\odot$ \citep{sawala_abundance_2013,sawala_chosen_2016,fitts_fire_2017,bose_imprint_2018}. We therefore include the effect of reionisation by setting $\epsilon_*(T_\mathrm{vir} < 10^4 \mathrm{K}) = 0$.

Of course, one could think of more complex ways in which the expected star formation efficiency might evolve with cosmic time. For instance, the evolution of cooling versus free-fall time of a cloud of gas, the evolution of metallicity and the UV background radiation, might all result in a more complex evolution. However, the aim here is to describe the main features of the universe as simply as possible, and so we leave exploration of more complex models for future work.

We calibrate the $\epsilon_*(T_\mathrm{vir})$ model to the GSMF at ${z}=0$ using the reduced chi-squared statistic to derive the best-fitting parameters. The best fit efficiency parameters are shown in \cref{tab:eps_mh}. 

It is important to highlight that the models were calibrated to reproduce only the observed GSMF at redshift $z \sim 0$. \Cref{fig:GSMF_fit} shows the best fit virial temperature model in blue. The figure shows that both the halo mass-dependent efficiency $\epsilon_*$, and the virial temperature-dependent efficiency $\epsilon_*(T_\mathrm{vir})$ models, provide a good fit to the present-day GSMF, both at the faint end and at the knee. \Cref{fig:GSMF_evol} shows that a star formation efficiency as a function of the virial temperature of the halo provides a good fit to the abundance of galaxies both at low and high redshift, but the evolution is too rapid at intermediate redshift ($z{=}1$ to $z{=}4$). 

\begin{figure}
\centering 
\includegraphics[width=0.48\textwidth]{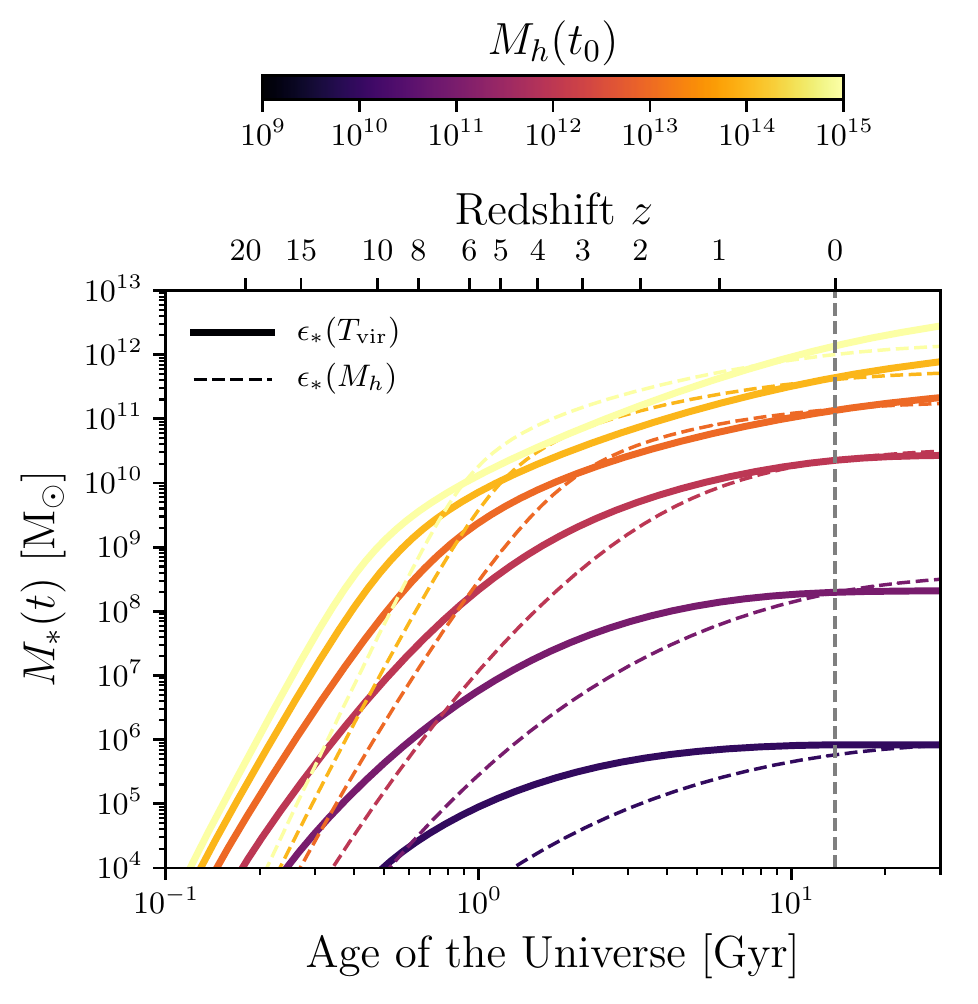}
  \vspace{-1.5em}
 \caption{Evolution of the stellar mass within haloes using the best-fit parameters for both models (\cref{tab:eps_mh}). The halo mass-dependent model is shown in dashed lines, while the virial temperature efficiency model is shown in solid lines. Colour coding represents different present-day halo masses $M_h(t_0)$. Once massive haloes reach the critical mass, the build up of stellar mass slows down significantly, i.e. the change in slope of the curves is due to AGN feedback becoming effective in those haloes.}
 \label{fig:M_star_evol}
\end{figure}

\begin{figure}
\centering 
\includegraphics[width=0.48\textwidth]{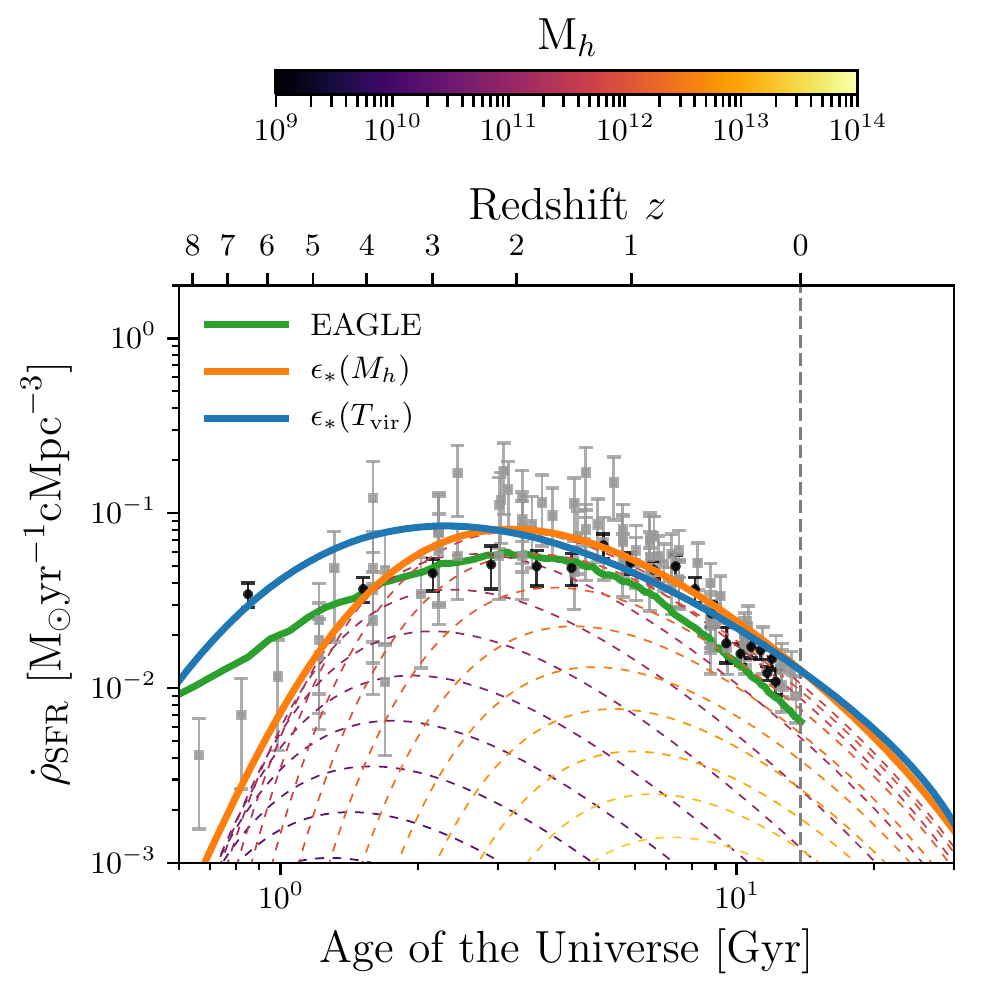}
  \vspace{-1.5em}
 \caption{The predicted SFR history of the Universe for the two efficiency models presented in this paper. Coloured lines show the contributions from dark matter haloes of different masses (per dex), using the star formation efficiency described by \cref{eq:ssfr_rate_density_dm}, and using the virial temperature efficiency model. The total SFR for the virial temperature efficiency model is shown in blue. The time-independent efficiency model is shown in orange. Results from the \textsc{eagle} simulation are shown in green for reference. Observational data compiled by \protect\cite{behroozi_average_2013} are shown as grey symbols. Observational data from \protect\cite{driver_gama/g10-cosmos/3d-hst:_2018} are shown as black symbols. The analytic model using a halo mass-dependent star formation efficiency reproduces the amplitude and shape of the cosmic SFR density remarkably well.}
 \label{fig:Madau}
\end{figure}
 
\Cref{fig:M_star_evol} shows the build up of the stellar mass within haloes using both models (calculated integrating \cref{eq:stellar_mass_int}). Colour coding represents different present-day halo mass $M_h(t_0)$. The transition of the star formation efficiency at $M_\mathrm{crit}$ can be clearly seen in very massive haloes, where there is a rapid rise of stellar mass, then, when the halo reaches the critical mass (or virial temperature), the build up of stellar mass slows down significantly. The change in slope is due to AGN feedback becoming efficient in those haloes, preventing any further star formation. 

\Cref{fig:Madau} shows the predicted cosmic SFR for the two efficiency models. Using the analytic model, we can clearly see the contribution to the integrated SFR density from dark matter haloes of different masses (per dex) shown as coloured dashed lines (only shown for the halo mass-dependent efficiency model). The total SFR for the virial temperature efficiency model is shown in blue. The halo mass-dependent efficiency model is shown in orange. Results from the \textsc{eagle} simulation are shown in green for reference. Observational data compiled by \cite{behroozi_average_2013} are shown as grey symbols. The latest observational results from the GAMA survey from \cite{driver_gama/g10-cosmos/3d-hst:_2018} are shown as black symbols. The model using a halo mass efficiency reproduces the amplitude and shape of the observed SFR density remarkably well, while the virial temperature-dependent efficiency model, produces a higher SFR at high redshift.

\begin{figure}
\centering 
\includegraphics[width=0.48\textwidth]{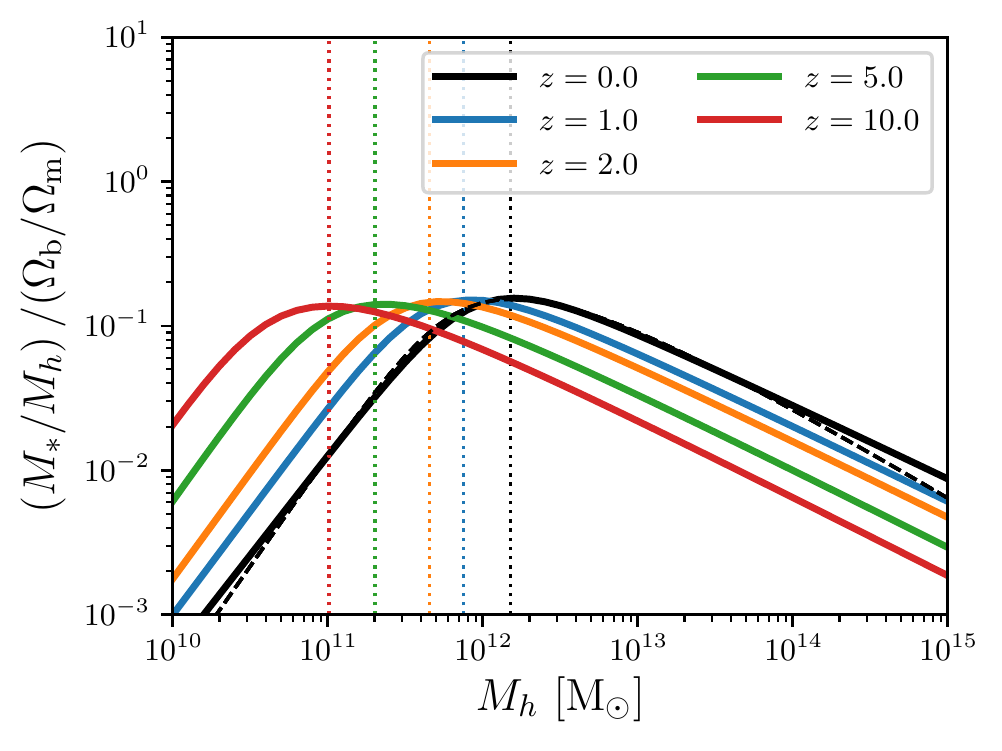}
  \vspace{-1.5em}
 \caption{Predicted SHMR for the time-independent, and evolving star formation efficiency models. Colour coding represents different observed redshifts. The halo mass-dependent model is shown as a dashed black line. Vertical dotted lines indicate the critical mass derived in \protect\cite{bower_dark_2017}, which tracks the triggering of a rapid black hole growth stage in the \textsc{eagle} simulations \protect\citep{mcalpine_rapid_2018}.}
 \label{fig:SHMR}
\end{figure}

\begin{figure*}
\centering 
\includegraphics[width=0.95\textwidth]{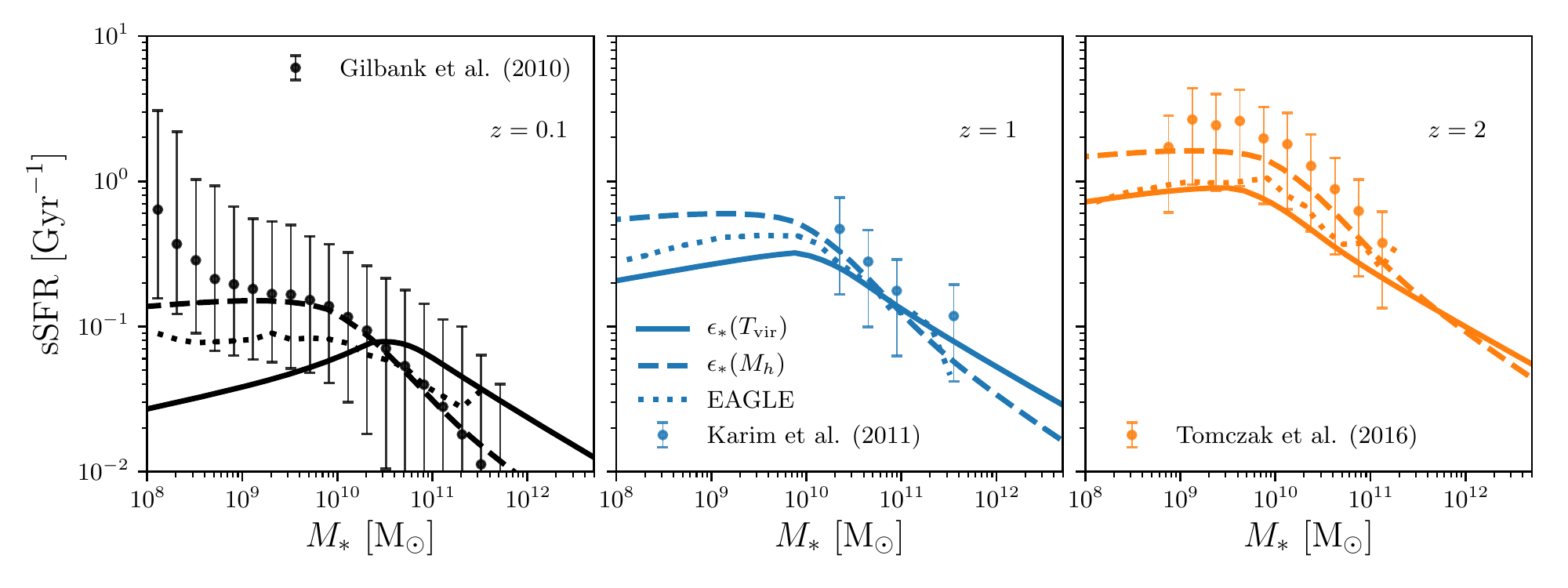}
  \vspace{-1.5em}
 \caption{The sSFR of galaxies at different redshifts. The model using a virial temperature efficiency is shown in solid lines. The halo mass-dependent model is shown in dashed lines. Results from the \textsc{eagle} simulations are shown in dotted lines for reference. Observational data from \protect\cite{gilbank_local_2010,karim_star_2011,tomczak_sfr-m*_2016} are shown as symbols. The halo mass-dependent model is in remarkable agreement with observational datasets.}
 \label{fig:sSFR}
\end{figure*}

\Cref{fig:SHMR} shows the predicted SHMR from both efficiency models. Colour coding represents different redshifts. The virial temperature efficiency model is shown in solid lines. The halo mass-dependent efficiency model is shown with a dashed line (only shown for ${z}=0$ as the halo mass-dependent efficiency model is constant in time). The critical halo mass predicted in \cite{bower_dark_2017} is shown in vertical dotted lines, which roughly coincide with the peak efficiency for a viral temperature efficiency model. Recently, \cite{mcalpine_rapid_2018} showed that the critical halo mass predicted in \cite{bower_dark_2017} agrees remarkably well with the triggering of a rapid black hole growth stage in the \textsc{eagle} simulations.

The model using a virial temperature efficiency predicts a SHMR relation that differs from observational contains using abundance and clustering properties of galaxy samples with predictions from a phenomenological halo models. For example, recently \cite{cowley_galaxyhalo_2018} calculated that the peak of the SHMR shifts to higher masses at earlier times. These methods however, depend heavily on the underlying modelling and assumptions. More sophisticated empirical models \citep[e.g.][]{behroozi_universemachine:_2018,moster_emerge_2018} find that the peak in the SHMR moves first to higher masses for low redshifts, and then to lower masses at high redshifts.  

Finally, in \cref{fig:sSFR} we show the sSFR of galaxies for different redshifts. The halo mass-dependent model is shown in dashed lines. The model using a virial temperature efficiency is shown in solid lines. Results for central galaxies from the \textsc{eagle} simulations are shown in dotted lines for reference. Observational data from \cite{gilbank_local_2010,karim_star_2011,tomczak_sfr-m*_2016} are shown as symbols. While not calibrated to reproduce the sSFR of galaxies, the agreement of the halo mass-dependent model with the observational data is remarkable.

\section{Discussion and Conclusions} \label{sec:con}

In our current paradigm of galaxy formation, every galaxy forms within a dark matter halo. Due to the tight correlation observed between the properties of galaxies and their host haloes, it is natural to expect that individual galaxy assembly could be correlated with halo assembly (see \citealt{wechsler_connection_2018} for a review).

In this paper we developed a fully analytic model of galaxy formation that connects the growth of dark matter haloes in a cosmological background, with the build up of stellar mass within these haloes. The model restricts the role of baryonic astrophysics to setting the relation between galaxies and their haloes. We assume an effective star formation efficiency which captures all the physical processes involved in the conversions of gas into stars, i.e. cooling, star formation law, feedback mechanisms, etc. 
 
We show that galaxy formation is revealed as a simple process where the effective star formation efficiency within haloes is only a function of their mass. We show that all the complex physics of galaxy formation, the interplay between cosmology and baryonic process can be understood as a simple set of equations. Despite its simplicity, the model reproduces self-consistently the shape and evolution of the cosmic star formation rate density, the specific star formation rate of galaxies, and the galaxy stellar mass function, both at the present time and at high redshift. 

We use our model to investigate the origin of the characteristic shape of the GSMF and the need for a double Schechter function to describe it. Using the  logarithmic slope of the SHMR, the model naturally explains an inflection point in the distribution causing the characteristic ``bump'' observed at the knee of the GSMF. 

 To demonstrate the flexibility and power of our mathematical framework, we introduced a physically motived model for the effective star formation efficiency, characterised by a time-independent critical virial temperature, $T_\mathrm{crit}$. The model assumes that there exists a critical halo virial temperature at which there is a transition from where star formation driven outflows can escape, to where outflows stall, triggering high star formation rates and rapid BH growth. We demonstrate that this model can reproduce the GSMF at high redshift ($z>4$) better than a simple halo mass-dependent model, but the evolution at intermediate redshifts is to rapid to reproduce observations. 

While the aim of this paper is not to present a ``perfect'' model fitted to reproduce a large set of observational constraints, the two variations of an effective star formation efficiency presented here, already provide very valuable information about the average evolution of the galaxy population within a cosmological background. Furthermore, the model can be easily extended to include further modelling (such as time evolution of the model parameters, or a prescription for satellite galaxies, e.g. \citealt{Grylls:2019}) or the use of advanced gradient-based minimization and Markov Chain Monte Carlo algorithms to fit to a larger number of datasets. Additionally, the model can be easily adapted to combine the equations developed here, with for example, halo merger trees from a dark matter simulation.

Our model is limited to the connection between haloes and central galaxies only. Sub-haloes and satellite galaxies are  subject to complex processes, such as tidal and ram pressure stripping, which are not included. 

Finally, one of the main advantages of the model is that by providing a set of analytic equations, the model can be easily ``inverted'' and allows for rapid experiments to be conducted, providing a great tool to explore the differential effects of baryonic physics, averaged over galaxy scales. We conclude therefore that there is a clear opportunity to use the analytic model developed in this paper to improve theoretical galaxy formation models.

\section*{Acknowledgements}
This project has received funding from the European Research Council (ERC) under the European Union’s Horizon 2020 research and innovation programme (grant agreement No 769130). This work was supported by the Science and Technology Facilities Council [ST/P000541/1].
This work used the DiRAC@Durham facility managed by the Institute for Computational Cosmology on behalf of the STFC DiRAC HPC Facility (www.dirac.ac.uk). The equipment was funded by BEIS capital funding via STFC capital grants ST/K00042X/1, ST/P002293/1 and ST/R002371/1, Durham University and STFC operations grant ST/R000832/1. DiRAC is part of the National e-Infrastructure.

%%%%%%%%%%%%%%%%%%%%%%%%%%%%%%%%%%%%%%%%%%%%%%%%%%



\bibliographystyle{mnras}
\bibliography{biblio} % if your bibtex file is called example.bib

\begin{thebibliography}{}
\makeatletter
\relax
\def\mn@urlcharsother{\let\do\@makeother \do\$\do\&\do\#\do\^\do\_\do\%\do\~}
\def\mn@doi{\begingroup\mn@urlcharsother \@ifnextchar [ {\mn@doi@}
  {\mn@doi@[]}}
\def\mn@doi@[#1]#2{\def\@tempa{#1}\ifx\@tempa\@empty \href
  {http://dx.doi.org/#2} {doi:#2}\else \href {http://dx.doi.org/#2} {#1}\fi
  \endgroup}
\def\mn@eprint#1#2{\mn@eprint@#1:#2::\@nil}
\def\mn@eprint@arXiv#1{\href {http://arxiv.org/abs/#1} {{\tt arXiv:#1}}}
\def\mn@eprint@dblp#1{\href {http://dblp.uni-trier.de/rec/bibtex/#1.xml}
  {dblp:#1}}
\def\mn@eprint@#1:#2:#3:#4\@nil{\def\@tempa {#1}\def\@tempb {#2}\def\@tempc
  {#3}\ifx \@tempc \@empty \let \@tempc \@tempb \let \@tempb \@tempa \fi \ifx
  \@tempb \@empty \def\@tempb {arXiv}\fi \@ifundefined
  {mn@eprint@\@tempb}{\@tempb:\@tempc}{\expandafter \expandafter \csname
  mn@eprint@\@tempb\endcsname \expandafter{\@tempc}}}

\bibitem[\protect\citeauthoryear{Angulo, Springel, White, Jenkins, Baugh  \&
  Frenk}{Angulo et~al.}{2012}]{angulo_scaling_2012}
Angulo R.~E.,  Springel V.,  White S. D.~M.,  Jenkins A.,  Baugh C.~M.,   Frenk
  C.~S.,  2012, \mn@doi [\mnras] {10.1111/j.1365-2966.2012.21830.x}, 426, 2046

\bibitem[\protect\citeauthoryear{Baldry, Glazebrook  \& Driver}{Baldry
  et~al.}{2008}]{baldry_galaxy_2008}
Baldry I.~K.,  Glazebrook K.,   Driver S.~P.,  2008, \mn@doi [Monthly Notices
  of the Royal Astronomical Society] {10.1111/j.1365-2966.2008.13348.x}, 388,
  945

\bibitem[\protect\citeauthoryear{Baldry et~al.,}{Baldry
  et~al.}{2012}]{baldry_galaxy_2012}
Baldry I.~K.,  et~al., 2012, \mn@doi [\mnras]
  {10.1111/j.1365-2966.2012.20340.x}, 421, 621

\bibitem[\protect\citeauthoryear{Behroozi, Wechsler  \& Conroy}{Behroozi
  et~al.}{2013}]{behroozi_average_2013}
Behroozi P.~S.,  Wechsler R.~H.,   Conroy C.,  2013, \mn@doi [\apj]
  {10.1088/0004-637X/770/1/57}, 770, 57

\bibitem[\protect\citeauthoryear{{Behroozi}, {Wechsler}, {Hearin}  \&
  {Conroy}}{{Behroozi} et~al.}{2019}]{behroozi_universemachine:_2018}
{Behroozi} P.,  {Wechsler} R.~H.,  {Hearin} A.~P.,   {Conroy} C.,  2019,
  \mn@doi [\mnras] {10.1093/mnras/stz1182}, \href
  {https://ui.adsabs.harvard.edu/abs/2019MNRAS.tmp.1134B} {p.~1134}

\bibitem[\protect\citeauthoryear{Benson}{Benson}{2012}]{benson_g_2012}
Benson A.~J.,  2012, \mn@doi [\na] {10.1016/j.newast.2011.07.004}, 17, 175

\bibitem[\protect\citeauthoryear{Benson, Bower, Frenk, Lacey, Baugh  \&
  Cole}{Benson et~al.}{2003}]{benson_what_2003}
Benson A.~J.,  Bower R.~G.,  Frenk C.~S.,  Lacey C.~G.,  Baugh C.~M.,   Cole
  S.,  2003, \mn@doi [\apj] {10.1086/379160}, 599, 38

\bibitem[\protect\citeauthoryear{Berlind \& Weinberg}{Berlind \&
  Weinberg}{2002}]{berlind_halo_2002}
Berlind A.~A.,  Weinberg D.~H.,  2002, \mn@doi [\apj] {10.1086/341469}, 575,
  587

\bibitem[\protect\citeauthoryear{{Bose}, {Deason}  \& {Frenk}}{{Bose}
  et~al.}{2018}]{bose_imprint_2018}
{Bose} S.,  {Deason} A.~J.,   {Frenk} C.~S.,  2018, \mn@doi [\apj]
  {10.3847/1538-4357/aacbc4}, \href
  {https://ui.adsabs.harvard.edu/abs/2018ApJ...863..123B} {863, 123}

\bibitem[\protect\citeauthoryear{Bouch{\'e} et~al.,}{Bouch{\'e}
  et~al.}{2010}]{bouche_impact_2010}
Bouch{\'e} N.,  et~al., 2010, \mn@doi [\apj] {10.1088/0004-637X/718/2/1001},
  718, 1001

\bibitem[\protect\citeauthoryear{Bower, Benson, Malbon, Helly, Frenk, Baugh,
  Cole  \& Lacey}{Bower et~al.}{2006}]{bower_breaking_2006}
Bower R.~G.,  Benson A.~J.,  Malbon R.,  Helly J.~C.,  Frenk C.~S.,  Baugh
  C.~M.,  Cole S.,   Lacey C.~G.,  2006, \mn@doi [\mnras]
  {10.1111/j.1365-2966.2006.10519.x}, 370, 645

\bibitem[\protect\citeauthoryear{Bower, Schaye, Frenk, Theuns, Schaller, Crain
  \& McAlpine}{Bower et~al.}{2017}]{bower_dark_2017}
Bower R.~G.,  Schaye J.,  Frenk C.~S.,  Theuns T.,  Schaller M.,  Crain R.~A.,
   McAlpine S.,  2017, \mn@doi [\mnras] {10.1093/mnras/stw2735}, 465, 32

\bibitem[\protect\citeauthoryear{Brainerd \& Specian}{Brainerd \&
  Specian}{2003}]{brainerd_mass--light_2003}
Brainerd T.~G.,  Specian M.~A.,  2003, \mn@doi [\apjl] {10.1086/378149}, 593,
  L7

\bibitem[\protect\citeauthoryear{Bruzual \& Charlot}{Bruzual \&
  Charlot}{2003}]{bruzual_stellar_2003}
Bruzual G.,  Charlot S.,  2003, \mn@doi [\mnras]
  {10.1046/j.1365-8711.2003.06897.x}, 344, 1000

\bibitem[\protect\citeauthoryear{Calzetti, Armus, Bohlin, Kinney, Koornneef  \&
  Storchi-Bergmann}{Calzetti et~al.}{2000}]{calzetti_dust_2000}
Calzetti D.,  Armus L.,  Bohlin R.~C.,  Kinney A.~L.,  Koornneef J.,
  Storchi-Bergmann T.,  2000, \mn@doi [\apj] {10.1086/308692}, 533, 682

\bibitem[\protect\citeauthoryear{Chabrier}{Chabrier}{2003}]{chabrier_galactic_2003}
Chabrier G.,  2003, \mn@doi [\pasp] {10.1086/376392}, 115, 763

\bibitem[\protect\citeauthoryear{Clauwens, Schaye, Franx  \& Bower}{Clauwens
  et~al.}{2018}]{clauwens_three_2018}
Clauwens B.,  Schaye J.,  Franx M.,   Bower R.~G.,  2018, \mn@doi [\mnras]
  {10.1093/mnras/sty1229}, 478, 3994

\bibitem[\protect\citeauthoryear{Cole, Aragon-Salamanca, Frenk, Navarro  \&
  Zepf}{Cole et~al.}{1994}]{cole_recipe_1994}
Cole S.,  Aragon-Salamanca A.,  Frenk C.~S.,  Navarro J.~F.,   Zepf S.~E.,
  1994, \mn@doi [\mnras] {10.1093/mnras/271.4.781}, 271, 781

\bibitem[\protect\citeauthoryear{Cooray \& Sheth}{Cooray \&
  Sheth}{2002}]{cooray_halo_2002}
Cooray A.,  Sheth R.,  2002, \mn@doi [\physrep]
  {10.1016/S0370-1573(02)00276-4}, 372, 1

\bibitem[\protect\citeauthoryear{Correa, Wyithe, Schaye  \& Duffy}{Correa
  et~al.}{2015}]{correa_accretion_2015}
Correa C.~A.,  Wyithe J. S.~B.,  Schaye J.,   Duffy A.~R.,  2015, \mn@doi
  [\mnras] {10.1093/mnras/stv689}, 450, 1514

\bibitem[\protect\citeauthoryear{Couchman \& Rees}{Couchman \&
  Rees}{1986}]{couchman_pregalactic_1986}
Couchman H. M.~P.,  Rees M.~J.,  1986, \mn@doi [\mnras]
  {10.1093/mnras/221.1.53}, 221, 53

\bibitem[\protect\citeauthoryear{Cowley et~al.,}{Cowley
  et~al.}{2018}]{cowley_galaxyhalo_2018}
Cowley W.~I.,  et~al., 2018, \mn@doi [\apj] {10.3847/1538-4357/aaa41d}, 853, 69

\bibitem[\protect\citeauthoryear{{Crain} et~al.,}{{Crain}
  et~al.}{2015}]{crain_eagle_2015}
{Crain} R.~A.,  et~al., 2015, \mn@doi [\mnras] {10.1093/mnras/stv725}, 450,
  1937

\bibitem[\protect\citeauthoryear{Dav{\'e}, Finlator  \& Oppenheimer}{Dav{\'e}
  et~al.}{2012a}]{dave_analytic_2012}
Dav{\'e} R.,  Finlator K.,   Oppenheimer B.~D.,  2012a, \mn@doi [\mnras]
  {10.1111/j.1365-2966.2011.20148.x}, 421, 98

\bibitem[\protect\citeauthoryear{{Dav{\'e}}, {Finlator}  \&
  {Oppenheimer}}{{Dav{\'e}} et~al.}{2012b}]{dave_11}
{Dav{\'e}} R.,  {Finlator} K.,   {Oppenheimer} B.~D.,  2012b, \mn@doi [\mnras]
  {10.1111/j.1365-2966.2011.20148.x}, \href
  {https://ui.adsabs.harvard.edu/abs/2012MNRAS.421...98D} {421, 98}

\bibitem[\protect\citeauthoryear{Dav{\'e}, Thompson  \& Hopkins}{Dav{\'e}
  et~al.}{2016}]{dave_mufasa:_2016}
Dav{\'e} R.,  Thompson R.,   Hopkins P.~F.,  2016, \mn@doi [\mnras]
  {10.1093/mnras/stw1862}, 462, 3265

\bibitem[\protect\citeauthoryear{{Dekel} \& {Silk}}{{Dekel} \&
  {Silk}}{1986}]{dekel_silk_86}
{Dekel} A.,  {Silk} J.,  1986, \mn@doi [\apj] {10.1086/164050}, \href
  {https://ui.adsabs.harvard.edu/abs/1986ApJ...303...39D} {303, 39}

\bibitem[\protect\citeauthoryear{Diemer, More  \& Kravtsov}{Diemer
  et~al.}{2013}]{diemer_pseudo-evolution_2013}
Diemer B.,  More S.,   Kravtsov A.~V.,  2013, \mn@doi [\apj]
  {10.1088/0004-637X/766/1/25}, 766, 25

\bibitem[\protect\citeauthoryear{Doroshkevich, Zel'dovich  \&
  Novikov}{Doroshkevich et~al.}{1967}]{doroshkevich_origin_1967}
Doroshkevich A.~G.,  Zel'dovich Y.~B.,   Novikov I.~D.,  1967, Soviet
  Astronomy, 11, 233

\bibitem[\protect\citeauthoryear{Driver et~al.,}{Driver
  et~al.}{2018}]{driver_gama/g10-cosmos/3d-hst:_2018}
Driver S.~P.,  et~al., 2018, \mn@doi [\mnras] {10.1093/mnras/stx2728}, 475,
  2891

\bibitem[\protect\citeauthoryear{{Dubois}, {Volonteri}, {Silk}, {Devriendt},
  {Slyz}  \& {Teyssier}}{{Dubois} et~al.}{2015}]{dubois_2015}
{Dubois} Y.,  {Volonteri} M.,  {Silk} J.,  {Devriendt} J.,  {Slyz} A.,
  {Teyssier} R.,  2015, \mn@doi [\mnras] {10.1093/mnras/stv1416}, \href
  {https://ui.adsabs.harvard.edu/abs/2015MNRAS.452.1502D} {452, 1502}

\bibitem[\protect\citeauthoryear{Dubois, Peirani, Pichon, Devriendt, Gavazzi,
  Welker  \& Volonteri}{Dubois et~al.}{2016}]{dubois_horizon-agn_2016}
Dubois Y.,  Peirani S.,  Pichon C.,  Devriendt J.,  Gavazzi R.,  Welker C.,
  Volonteri M.,  2016, \mn@doi [\mnras] {10.1093/mnras/stw2265}, 463, 3948

\bibitem[\protect\citeauthoryear{Efstathiou}{Efstathiou}{1992}]{efstathiou_suppressing_1992}
Efstathiou G.,  1992, \mn@doi [\mnras] {10.1093/mnras/256.1.43P}, 256, 43P

\bibitem[\protect\citeauthoryear{Finlator \& Dav{\'e}}{Finlator \&
  Dav{\'e}}{2008}]{finlator_origin_2008}
Finlator K.,  Dav{\'e} R.,  2008, \mn@doi [\mnras]
  {10.1111/j.1365-2966.2008.12991.x}, 385, 2181

\bibitem[\protect\citeauthoryear{Fitts et~al.,}{Fitts
  et~al.}{2017}]{fitts_fire_2017}
Fitts A.,  et~al., 2017, \mn@doi [\mnras] {10.1093/mnras/stx1757}, 471, 3547

\bibitem[\protect\citeauthoryear{Fosalba, Crocce, Gazta{\~n}aga  \&
  Castander}{Fosalba et~al.}{2015}]{fosalba_mice_2015}
Fosalba P.,  Crocce M.,  Gazta{\~n}aga E.,   Castander F.~J.,  2015, \mn@doi
  [\mnras] {10.1093/mnras/stv138}, 448, 2987

\bibitem[\protect\citeauthoryear{Gilbank, Baldry, Balogh, Glazebrook  \&
  Bower}{Gilbank et~al.}{2010}]{gilbank_local_2010}
Gilbank D.~G.,  Baldry I.~K.,  Balogh M.~L.,  Glazebrook K.,   Bower R.~G.,
  2010, \mn@doi [\mnras] {10.1111/j.1365-2966.2010.16640.x}, 405, 2594

\bibitem[\protect\citeauthoryear{{Grylls}, {Shankar}, {Zanisi}  \&
  {Bernardi}}{{Grylls} et~al.}{2019}]{Grylls:2019}
{Grylls} P.~J.,  {Shankar} F.,  {Zanisi} L.,   {Bernardi} M.,  2019, \mn@doi
  [\mnras] {10.1093/mnras/sty3281}, \href
  {https://ui.adsabs.harvard.edu/abs/2019MNRAS.483.2506G} {483, 2506}

\bibitem[\protect\citeauthoryear{Haas, Schaye, Booth, Dalla~Vecchia, Springel,
  Theuns  \& Wiersma}{Haas et~al.}{2013}]{haas_physical_2013}
Haas M.~R.,  Schaye J.,  Booth C.~M.,  Dalla~Vecchia C.,  Springel V.,  Theuns
  T.,   Wiersma R. P.~C.,  2013, \mn@doi [\mnras] {10.1093/mnras/stt1487}, 435,
  2931

\bibitem[\protect\citeauthoryear{Henriques, White, Thomas, Angulo, Guo, Lemson,
  Springel  \& Overzier}{Henriques et~al.}{2015}]{henriques_galaxy_2015}
Henriques B. M.~B.,  White S. D.~M.,  Thomas P.~A.,  Angulo R.,  Guo Q.,
  Lemson G.,  Springel V.,   Overzier R.,  2015, \mn@doi [\mnras]
  {10.1093/mnras/stv705}, 451, 2663

\bibitem[\protect\citeauthoryear{Hoekstra, Yee  \& Gladders}{Hoekstra
  et~al.}{2004}]{hoekstra_properties_2004}
Hoekstra H.,  Yee H. K.~C.,   Gladders M.~D.,  2004, \mn@doi [\apj]
  {10.1086/382726}, 606, 67

\bibitem[\protect\citeauthoryear{Hudson et~al.,}{Hudson
  et~al.}{2015}]{hudson_cfhtlens:_2015}
Hudson M.~J.,  et~al., 2015, \mn@doi [\mnras] {10.1093/mnras/stu2367}, 447, 298

\bibitem[\protect\citeauthoryear{Ilbert et~al.,}{Ilbert
  et~al.}{2013}]{ilbert_mass_2013}
Ilbert O.,  et~al., 2013, \mn@doi [\aap] {10.1051/0004-6361/201321100}, 556,
  A55

\bibitem[\protect\citeauthoryear{Karim et~al.,}{Karim
  et~al.}{2011}]{karim_star_2011}
Karim A.,  et~al., 2011, \mn@doi [\apj] {10.1088/0004-637X/730/2/61}, 730, 61

\bibitem[\protect\citeauthoryear{Klypin, Trujillo-Gomez  \& Primack}{Klypin
  et~al.}{2011}]{klypin_dark_2011}
Klypin A.~A.,  Trujillo-Gomez S.,   Primack J.,  2011, \mn@doi [\apj]
  {10.1088/0004-637X/740/2/102}, 740, 102

\bibitem[\protect\citeauthoryear{Lacey et~al.,}{Lacey
  et~al.}{2016}]{lacey_unified_2016}
Lacey C.~G.,  et~al., 2016, \mn@doi [\mnras] {10.1093/mnras/stw1888}, 462, 3854

\bibitem[\protect\citeauthoryear{Lan, M{\'e}nard  \& Mo}{Lan
  et~al.}{2016}]{lan_galaxy_2016}
Lan T.-W.,  M{\'e}nard B.,   Mo H.,  2016, \mn@doi [Monthly Notices of the
  Royal Astronomical Society] {10.1093/mnras/stw898}, 459, 3998

\bibitem[\protect\citeauthoryear{Li \& White}{Li \&
  White}{2009}]{li_distribution_2009}
Li C.,  White S. D.~M.,  2009, \mn@doi [\mnras]
  {10.1111/j.1365-2966.2009.15268.x}, 398, 2177

\bibitem[\protect\citeauthoryear{Loeb \& Barkana}{Loeb \&
  Barkana}{2001}]{loeb_reionization_2001}
Loeb A.,  Barkana R.,  2001, \mn@doi [Annual Review of Astronomy and
  Astrophysics] {10.1146/annurev.astro.39.1.19}, 39, 19

\bibitem[\protect\citeauthoryear{Madau \& Dickinson}{Madau \&
  Dickinson}{2014}]{madau_cosmic_2014}
Madau P.,  Dickinson M.,  2014, \mn@doi [\araa]
  {10.1146/annurev-astro-081811-125615}, 52, 415

\bibitem[\protect\citeauthoryear{McAlpine et~al.,}{McAlpine
  et~al.}{2016}]{mcalpine_eagle_2016}
McAlpine S.,  et~al., 2016, \mn@doi [Astronomy and Computing]
  {10.1016/j.ascom.2016.02.004}, 15, 72

\bibitem[\protect\citeauthoryear{{McAlpine}, {Bower}, {Rosario}, {Crain},
  {Schaye}  \& {Theuns}}{{McAlpine} et~al.}{2018}]{mcalpine_rapid_2018}
{McAlpine} S.,  {Bower} R.~G.,  {Rosario} D.~J.,  {Crain} R.~A.,  {Schaye} J.,
   {Theuns} T.,  2018, \mn@doi [\mnras] {10.1093/mnras/sty2489}, \href
  {https://ui.adsabs.harvard.edu/abs/2018MNRAS.481.3118M} {481, 3118}

\bibitem[\protect\citeauthoryear{Mitchell, Lacey, Baugh  \& Cole}{Mitchell
  et~al.}{2016}]{mitchell_evolution_2016}
Mitchell P.~D.,  Lacey C.~G.,  Baugh C.~M.,   Cole S.,  2016, \mn@doi [\mnras]
  {10.1093/mnras/stv2741}, 456, 1459

\bibitem[\protect\citeauthoryear{Moster, Somerville, Maulbetsch, Bosch,
  Macci{\`o}, Naab  \& Oser}{Moster et~al.}{2010}]{moster_constraints_2010}
Moster B.~P.,  Somerville R.~S.,  Maulbetsch C.,  Bosch F. C. v.~d.,
  Macci{\`o} A.~V.,  Naab T.,   Oser L.,  2010, \mn@doi [\apj]
  {10.1088/0004-637X/710/2/903}, 710, 903

\bibitem[\protect\citeauthoryear{Moster, Naab  \& White}{Moster
  et~al.}{2013}]{moster_galactic_2013}
Moster B.~P.,  Naab T.,   White S. D.~M.,  2013, \mn@doi [\mnras]
  {10.1093/mnras/sts261}, 428, 3121

\bibitem[\protect\citeauthoryear{Moster, Naab  \& White}{Moster
  et~al.}{2018}]{moster_emerge_2018}
Moster B.~P.,  Naab T.,   White S. D.~M.,  2018, \mn@doi [\mnras]
  {10.1093/mnras/sty655}, 477, 1822

\bibitem[\protect\citeauthoryear{Moustakas et~al.,}{Moustakas
  et~al.}{2013}]{moustakas_primus:_2013}
Moustakas J.,  et~al., 2013, \mn@doi [\apj] {10.1088/0004-637X/767/1/50}, 767,
  50

\bibitem[\protect\citeauthoryear{Muzzin et~al.,}{Muzzin
  et~al.}{2013}]{muzzin_evolution_2013}
Muzzin A.,  et~al., 2013, \mn@doi [\apj] {10.1088/0004-637X/777/1/18}, 777, 18

\bibitem[\protect\citeauthoryear{Naab \& Ostriker}{Naab \&
  Ostriker}{2017}]{naab_theoretical_2017}
Naab T.,  Ostriker J.~P.,  2017, \mn@doi [\araa]
  {10.1146/annurev-astro-081913-040019}, 55, 59

\bibitem[\protect\citeauthoryear{Neistein, van~den Bosch  \& Dekel}{Neistein
  et~al.}{2006}]{neistein_natural_2006}
Neistein E.,  van~den Bosch F.~C.,   Dekel A.,  2006, \mn@doi [\mnras]
  {10.1111/j.1365-2966.2006.10918.x}, 372, 933

\bibitem[\protect\citeauthoryear{Neyman \& Scott}{Neyman \&
  Scott}{1952}]{neyman_theory_1952}
Neyman J.,  Scott E.~L.,  1952, \mn@doi [\apj] {10.1086/145599}, 116, 144

\bibitem[\protect\citeauthoryear{Norberg, Frenk  \& Cole}{Norberg
  et~al.}{2008}]{norberg_massive_2008}
Norberg P.,  Frenk C.~S.,   Cole S.,  2008, \mn@doi [\mnras]
  {10.1111/j.1365-2966.2007.12583.x}, 383, 646

\bibitem[\protect\citeauthoryear{Pillepich et~al.,}{Pillepich
  et~al.}{2018a}]{pillepich_simulating_2018}
Pillepich A.,  et~al., 2018a, \mn@doi [\mnras] {10.1093/mnras/stx2656}, 473,
  4077

\bibitem[\protect\citeauthoryear{{Pillepich} et~al.,}{{Pillepich}
  et~al.}{2018b}]{pillepich_first_results_2018}
{Pillepich} A.,  et~al., 2018b, \mn@doi [\mnras] {10.1093/mnras/stx3112}, \href
  {http://adsabs.harvard.edu/abs/2018MNRAS.475..648P} {475, 648}

\bibitem[\protect\citeauthoryear{{Planck Collaboration} et~al.,}{{Planck
  Collaboration} et~al.}{2014}]{planck_collaboration_planck_2014}
{Planck Collaboration} et~al., 2014, \mn@doi [\aap]
  {10.1051/0004-6361/201321591}, 571, A16

\bibitem[\protect\citeauthoryear{{Planck Collaboration} et~al.,}{{Planck
  Collaboration} et~al.}{2016}]{planck_collaboration_planck_2016}
{Planck Collaboration} et~al., 2016, \mn@doi [\aap]
  {10.1051/0004-6361/201525830}, 594, A13

\bibitem[\protect\citeauthoryear{Press \& Schechter}{Press \&
  Schechter}{1974}]{press_formation_1974}
Press W.~H.,  Schechter P.,  1974, \mn@doi [\apj] {10.1086/152650}, 187, 425

\bibitem[\protect\citeauthoryear{Qu et~al.,}{Qu
  et~al.}{2017}]{qu_chronicle_2017}
Qu Y.,  et~al., 2017, \mn@doi [\mnras] {10.1093/mnras/stw2437}, 464, 1659

\bibitem[\protect\citeauthoryear{Rees}{Rees}{1986}]{rees_lyman_1986}
Rees M.~J.,  1986, \mn@doi [\mnras] {10.1093/mnras/218.1.25P}, 218, 25P

\bibitem[\protect\citeauthoryear{{Rodriguez-Gomez} et~al.,}{{Rodriguez-Gomez}
  et~al.}{2016}]{Rodriguez_Gomez_2016}
{Rodriguez-Gomez} V.,  et~al., 2016, \mn@doi [\mnras] {10.1093/mnras/stw456},
  \href {http://adsabs.harvard.edu/abs/2016MNRAS.458.2371R} {458, 2371}

\bibitem[\protect\citeauthoryear{Rodr{\'i}guez-Puebla, Primack, Behroozi  \&
  Faber}{Rodr{\'i}guez-Puebla et~al.}{2016}]{rodriguez-puebla_is_2016}
Rodr{\'i}guez-Puebla A.,  Primack J.~R.,  Behroozi P.,   Faber S.~M.,  2016,
  \mn@doi [\mnras] {10.1093/mnras/stv2513}, 455, 2592

\bibitem[\protect\citeauthoryear{Salcido et~al.,}{Salcido
  et~al.}{2018}]{salcido_impact_2018}
Salcido J.,  et~al., 2018, \mn@doi [\mnras] {10.1093/mnras/sty879}, 477, 3744

\bibitem[\protect\citeauthoryear{Sawala, Frenk, Crain, Jenkins, Schaye, Theuns
  \& Zavala}{Sawala et~al.}{2013}]{sawala_abundance_2013}
Sawala T.,  Frenk C.~S.,  Crain R.~A.,  Jenkins A.,  Schaye J.,  Theuns T.,
  Zavala J.,  2013, \mn@doi [\mnras] {10.1093/mnras/stt259}, 431, 1366

\bibitem[\protect\citeauthoryear{Sawala et~al.,}{Sawala
  et~al.}{2016}]{sawala_chosen_2016}
Sawala T.,  et~al., 2016, \mn@doi [\mnras] {10.1093/mnras/stv2597}, 456, 85

\bibitem[\protect\citeauthoryear{Schaye et~al.,}{Schaye
  et~al.}{2010}]{schaye_physics_2010}
Schaye J.,  et~al., 2010, \mn@doi [\mnras] {10.1111/j.1365-2966.2009.16029.x},
  402, 1536

\bibitem[\protect\citeauthoryear{Schaye et~al.,}{Schaye
  et~al.}{2015}]{schaye_eagle_2015}
Schaye J.,  et~al., 2015, \mn@doi [\mnras] {10.1093/mnras/stu2058}, 446, 521

\bibitem[\protect\citeauthoryear{Schechter}{Schechter}{1976}]{schechter_analytic_1976}
Schechter P.,  1976, \mn@doi [The Astrophysical Journal] {10.1086/154079}, 203,
  297

\bibitem[\protect\citeauthoryear{{Schmidt}}{{Schmidt}}{1963}]{schmidt_rate_1963}
{Schmidt} M.,  1963, \mn@doi [\apj] {10.1086/147553}, \href
  {http://adsabs.harvard.edu/abs/1963ApJ...137..758S} {137, 758}

\bibitem[\protect\citeauthoryear{Sharma \& Theuns}{Sharma \&
  Theuns}{2019}]{sharma_2019}
Sharma M.,  Theuns T.,  2019, arXiv e-prints, p. arXiv:1906.10135

\bibitem[\protect\citeauthoryear{Somerville \& Dav{\'e}}{Somerville \&
  Dav{\'e}}{2015}]{somerville_physical_2015}
Somerville R.~S.,  Dav{\'e} R.,  2015, \mn@doi [\araa]
  {10.1146/annurev-astro-082812-140951}, 53, 51

\bibitem[\protect\citeauthoryear{Somerville, Hopkins, Cox, Robertson  \&
  Hernquist}{Somerville et~al.}{2008}]{somerville_semi-analytic_2008}
Somerville R.~S.,  Hopkins P.~F.,  Cox T.~J.,  Robertson B.~E.,   Hernquist L.,
   2008, \mn@doi [\mnras] {10.1111/j.1365-2966.2008.13805.x}, 391, 481

\bibitem[\protect\citeauthoryear{Song et~al.,}{Song
  et~al.}{2016}]{song_evolution_2016}
Song M.,  et~al., 2016, \mn@doi [\apj] {10.3847/0004-637X/825/1/5}, 825, 5

\bibitem[\protect\citeauthoryear{Springel et~al.,}{Springel
  et~al.}{2005}]{springel_simulations_2005}
Springel V.,  et~al., 2005, \mn@doi [\nat] {10.1038/nature03597}, 435, 629

\bibitem[\protect\citeauthoryear{Tacchella, Bose, Conroy, Eisenstein  \&
  Johnson}{Tacchella et~al.}{2018}]{tacchella_redshift-independent_2018}
Tacchella S.,  Bose S.,  Conroy C.,  Eisenstein D.~J.,   Johnson B.~D.,  2018,
  preprint

\bibitem[\protect\citeauthoryear{Tomczak et~al.,}{Tomczak
  et~al.}{2014}]{tomczak_galaxy_2014}
Tomczak A.~R.,  et~al., 2014, \mn@doi [\apj] {10.1088/0004-637X/783/2/85}, 783,
  85

\bibitem[\protect\citeauthoryear{Tomczak et~al.,}{Tomczak
  et~al.}{2016}]{tomczak_sfr-m*_2016}
Tomczak A.~R.,  et~al., 2016, \mn@doi [\apj] {10.3847/0004-637X/817/2/118},
  817, 118

\bibitem[\protect\citeauthoryear{Trujillo-Gomez, Klypin, Primack  \&
  Romanowsky}{Trujillo-Gomez et~al.}{2011}]{trujillo-gomez_galaxies_2011}
Trujillo-Gomez S.,  Klypin A.,  Primack J.,   Romanowsky A.~J.,  2011, \mn@doi
  [\apj] {10.1088/0004-637X/742/1/16}, 742, 16

\bibitem[\protect\citeauthoryear{Vogelsberger et~al.,}{Vogelsberger
  et~al.}{2014}]{vogelsberger_introducing_2014}
Vogelsberger M.,  et~al., 2014, \mn@doi [\mnras] {10.1093/mnras/stu1536}, 444,
  1518

\bibitem[\protect\citeauthoryear{Wechsler \& Tinker}{Wechsler \&
  Tinker}{2018}]{wechsler_connection_2018}
Wechsler R.~H.,  Tinker J.~L.,  2018, preprint

\bibitem[\protect\citeauthoryear{Wetzel \& Nagai}{Wetzel \&
  Nagai}{2015}]{wetzel_physical_2015}
Wetzel A.~R.,  Nagai D.,  2015, \mn@doi [\apj] {10.1088/0004-637X/808/1/40},
  808, 40

\bibitem[\protect\citeauthoryear{White \& Frenk}{White \&
  Frenk}{1991}]{white_galaxy_1991}
White S. D.~M.,  Frenk C.~S.,  1991, \mn@doi [\apj] {10.1086/170483}, 379, 52

\bibitem[\protect\citeauthoryear{{White} \& {Rees}}{{White} \&
  {Rees}}{1978}]{white_rees78}
{White} S.~D.~M.,  {Rees} M.~J.,  1978, \mn@doi [\mnras]
  {10.1093/mnras/183.3.341}, \href
  {https://ui.adsabs.harvard.edu/abs/1978MNRAS.183..341W} {183, 341}

\bibitem[\protect\citeauthoryear{Yang, Mo  \& van~den Bosch}{Yang
  et~al.}{2009}]{yang_galaxy_2009}
Yang X.,  Mo H.~J.,   van~den Bosch F.~C.,  2009, \mn@doi [The Astrophysical
  Journal] {10.1088/0004-637X/695/2/900}, 695, 900

\bibitem[\protect\citeauthoryear{Yang, Mo, van~den Bosch, Zhang  \& Han}{Yang
  et~al.}{2012}]{yang_evolution_2012}
Yang X.,  Mo H.~J.,  van~den Bosch F.~C.,  Zhang Y.,   Han J.,  2012, \mn@doi
  [\apj] {10.1088/0004-637X/752/1/41}, 752, 41

\bibitem[\protect\citeauthoryear{Zaritsky, Smith, Frenk  \& White}{Zaritsky
  et~al.}{1993}]{zaritsky_satellites_1993}
Zaritsky D.,  Smith R.,  Frenk C.,   White S. D.~M.,  1993, \mn@doi [\apj]
  {10.1086/172379}, 405, 464

\bibitem[\protect\citeauthoryear{van~den Bosch, Norberg, Mo  \& Yang}{van~den
  Bosch et~al.}{2004}]{van_den_bosch_probing_2004}
van~den Bosch F.~C.,  Norberg P.,  Mo H.~J.,   Yang X.,  2004, \mn@doi [\mnras]
  {10.1111/j.1365-2966.2004.08021.x}, 352, 1302

\bibitem[\protect\citeauthoryear{van~den Bosch, Ogiya, Hahn  \&
  Burkert}{van~den Bosch et~al.}{2018}]{van_den_bosch_disruption_2018}
van~den Bosch F.~C.,  Ogiya G.,  Hahn O.,   Burkert A.,  2018, \mn@doi [\mnras]
  {10.1093/mnras/stx2956}, 474, 3043

\makeatother
\end{thebibliography}

%\begin{thebibliography}{99}
%\end{thebibliography}

%%%%%%%%%%%%%%%%%%%%%%%%%%%%%%%%%%%%%%%%%%%%%%%%%%

%%%%%%%%%%%%%%%%% APPENDICES %%%%%%%%%%%%%%%%%%%%%

\appendix
\renewcommand{\thetable}{\thesection\arabic{table}}
\renewcommand{\thefigure}{\thesection\arabic{figure}} 

\section{Derivation of the Model}\label{sec:derivation}

\subsection{Cosmological expansion}

Here we provide a brief summary of the analytic solution of the Friedmann equation developed in \cite{salcido_impact_2018}. Equations with full cosmology dependence of the numerical constants are highlighted using a coloured superscript $(^{\textcolor{red}{c^*}})$, and can be used for arbitrary flat $\Lambda$CDM cosmologies. 

Using a Taylor expansion, the expansion factor of the Universe can be written as,
\begin{equation}\label{A:eq:a_of_t_power}
a^{\textcolor{red}{c^*}}(t) \propto \left[\frac{3}{2} \frac{t}{t_m}\right]^{2/3}\left(1 + \frac{1}{4} \left(\frac{t}{t_\Lambda}\right)^2 + \frac{1}{80} \left(\frac{t}{t_\Lambda}\right)^4 +...\right),
\end{equation}
where the \textit{matter timescale} is given by,
 \begin{equation}\label{A:eq:tm}
    t^{\textcolor{red}{c^*}}_\mathrm{m} = \sqrt{\frac{3}{8\pi G{\rho}_0}} = \frac{1}{{H}_0 \sqrt{{\Omega}_{\mathrm{m},0}}},
 \end{equation}
 and the \textit{dark energy timescale} is given by,
  \begin{equation}\label{A:eq:tlambda}
    t^{\textcolor{red}{c^*}}_\Lambda = \sqrt{\frac{3}{\Lambda c^2}} = \frac{1}{H_0 \sqrt{\Omega_{\Lambda,0}}}.
 \end{equation}
For the cosmological parameters given at the end of Section~1,
$t_\mathrm{m} = 26.04 \Gyr$ and $t_\Lambda = 17.33 \Gyr$. At the present day, $t \equiv t_0 = 13.82 \Gyr $, so that $\frac{t_0}{t_m}=0.53$ and $\frac{t_0}{t_\Lambda} = 0.8$.
By convention, \cref{A:eq:a_of_t_power} is normalised so that $a(t_0) = 1$.

\subsection{The growth of density perturbations and the halo accretion rates}

Dark matter structures are assumed to have grown from small initial density perturbations. Expressing the density, $\rho$, in terms of the density perturbation contrast against a density background,
\begin{equation}
    \rho(\mathbf{x},t) = \bar{\rho}(t)[1+\delta(\mathbf{x},t)],
\end{equation}
the differential equation that governs the time dependence of the growth of linear perturbations in a pressureless fluid, such as e.g. dark matter, can be written as
\begin{equation}\label{A:eq:perturbations}
    \frac{\dd^2 \delta}{\dd t^2} + 2 \frac{\dot{a}}{a} \frac{\dd \delta}{\dd t} - 4\pi G \bar{\rho} \delta = 0.
\end{equation}

The growing mode of \cref{A:eq:perturbations} can be written as,
\begin{equation}
    \delta(t) = {D(t)} \delta(t_0),
\end{equation}
where $D(t)$ is the linear growth factor, which determines the normalisation of the linear matter power spectrum relative to the initial density perturbation power spectrum, and is computed by the integral 
\begin{equation}\label{A:eq:D_t}
D^{\textcolor{red}{c^*}}(t) \propto \frac{\dot{a}}{a} \int_0^t \frac{\dd t^\prime}{\dot{a}^2(t^\prime)}.
\end{equation}
Using the power-series approximation for $a(t)$ from \cref{A:eq:a_of_t_power}, keeping the leading order terms and using the definition of $f_{\Lambda}$ in \cref{eq:f_lambda}, we can obtain an analytic solution of \cref{A:eq:D_t},
\small  
\begin{equation}\label{A:eq:D_of_t_approx}
D^{\color{red}c^*}(t) = \left[\frac{3}{2} \frac{t}{t_m}\right]^{2/3} \frac{2}{5} t_m^{2} K_D \, f_\Lambda(t, - 0.16, 0.04), 
\end{equation}
\normalsize
where $K_D$ is a normalisation constant with units of time$^{-2}$. By convention, $K_D$ is chosen so that ${D}(t_0) = 1$. For the cosmological parameters inferred by the \citet{planck_collaboration_planck_2014}, $K_D = 4.7 \times 10^{-3} \, \mathrm{Gyr}^{-2}$. Collecting the numerical and cosmology dependent constants together gives, 
\small  
\begin{equation}\label{A:eq:D_of_t_approx_consts}
D(t) \approx 1.671 \left[\frac{t}{t_m}\right]^{2/3} f_\Lambda(t, - 0.16, 0.04). 
\end{equation}
\normalsize

The growth rates of linear perturbations do not directly predict the growth rates of haloes; however, we can directly connect the two through the approach developed by \cite{press_formation_1974}. \cite{correa_accretion_2015} showed that the accretion rates of haloes can be written as \citep[see also][]{neistein_natural_2006}, 
\begin{equation}\label{A:eq:mdot_eps}
\left(\frac{1}{M_h} \frac{\dd M_h}{\dd t}\right)^{\textcolor{red}{c^*}} = \sqrt{\frac{2}{\pi}}
     \frac{(\delta_c/D)}{S^{1/2}\left(q^\gamma-1\right)^{1/2}}  
     \frac{1}{D}\frac{\dd D}{\dd t},
\end{equation}
where $M_h$ is the halo mass and $S$ is the variance of the density field on the length scale corresponding the halo mass. $\delta_c$ is a parameter that represents a threshold in the linearly extrapolated density field for halo collapse. We assume $\delta_c=1.68$ \citep{press_formation_1974}. The parameters, $q$ and $\gamma$, are related to the shape of the power-spectrum around the halo mass $M_h$. The scale dependence of the density field is approximated as a power-law around $10^{12} \Msol$ haloes as $S=S_0 (M_h/10^{12} \mathrm{M}_\odot)^{-\gamma}$. \cite{correa_accretion_2015} find that this prescription works for different cosmologies because the halo mass histories are mainly driven by changes in $\sigma_{8}$ and $\Omega_{m}$. For the cosmological parameters inferred by the \citet{planck_collaboration_planck_2014}, $S_0\approx 3.98, \gamma\approx 0.3$ and $q\approx 3.16$. Collecting the numerical and cosmology dependent constants together yields, 
\begin{equation}\label{A:eq:mdot_eps_consts}
\frac{1}{M_h} \frac{\dd M_h}{\dd t} = 1.05
     \left(\frac{M_h}{10^{12}\mathrm{M}_\odot}\right)^{-\gamma/2}
     \frac{1}{D^2}\frac{\dd D}{\dd t}.
\end{equation}

Using the series approximation \cref{A:eq:D_of_t_approx_consts}, the specific growth rate of haloes can be written as,
\small
\begin{equation}\label{A:eq:mdot_eps_series}
\begin{aligned}
    \left(\frac{1}{M_h} \frac{\dd M_h}{\dd t} \right) ^{\color{red}c^*} = \frac{2.66}{\sqrt{S_0}K_D \, t_m^{3}} & \left(\frac{t}{t_m}\right)^{-5/3}
    \left(\frac{M_h}{10^{12}\mathrm{M}_\odot}\right)^{\gamma/2} f_\Lambda(t,-0.32,0.06).
\end{aligned}
\end{equation}
\normalsize

This differential equation can be solved by separation of variables to obtain the average mass history of dark matter haloes, 
\small
\begin{equation}
    \begin{aligned}
        \int_{M}^{M_0} & \left(\frac{M_h^\prime}{10^{12}\mathrm{M}_\odot}\right)^{-\left(\frac{\gamma}{2} +1\right)} \frac{\dd M_h}{10^{12}\mathrm{M}_\odot} \\ 
        = & \frac{2.66}{\sqrt{S_0}K_D \, t_m^{3}} 
           \int_{t}^{t_0} \left(\frac{t^\prime}{t_m}\right)^{-5/3} f_\Lambda(t,-0.32,0.06) \, {\dd t^\prime}
    \end{aligned}
\end{equation}
\normalsize
where $M_0$ is the mass of a halo today. Integrating both sides and solving for $M(t)$ yields, 
\small
\begin{equation}\label{A:eq:halo_mass_implicit}
    \begin{aligned}
 & \frac{2}{\gamma} \left[\left(\frac{M_h}{10^{12}\mathrm{M}_\odot} \right)^{-\gamma/2} 
   - \left(\frac{M_0}{10^{12}\mathrm{M}_\odot}\right)^{-\gamma/2}\right] =  \\  
   & \frac{4}{\sqrt{S_0}K_D t_m^{2}} \,
    \left[\left(\frac{t}{t_m}\right)^{-2/3}f_\Lambda(t, 0.16, - 0.01) \, - \left(\frac{t_0}{t_m}\right)^{-2/3}f_\Lambda(t_0, 0.16, - 0.01)\right].
    \end{aligned}
\end{equation}
\normalsize
Note that in the case $\gamma\rightarrow0$ the LHS becomes the logarithm of the mass ratio $M_h/M_0$, and all haloes grow by the same factor in a given time interval. For realistic power spectra, however, the relative growth rate increases with mass because massive haloes arise from increasingly rare fluctuations in the initial density perturbation field. We can re-write \Cref{A:eq:halo_mass_implicit} as an explicit equation for the halo mass as a function of time. This form is useful for symbolic substitution into calculations that are driven by the halo mass. 
\small
\begin{equation}\label{A:eq:halo_mass}
\begin{aligned}
& \frac{M_h^{\textcolor{red}{c^*}}(t)}{10^{12}\mathrm{M}_\odot} =  \left\{\left(\frac{M_0}{10^{12}\mathrm{M}_\odot}\right)^{-\gamma /2} + \right. \\
& \left. \frac{2 \gamma}{\sqrt{S_0}K_D t_m^{2}}\, 
    \left[\left(\frac{t}{t_m}\right)^{-2/3}f_\Lambda(t, 0.16, - 0.01) - 
         \left(\frac{t_0}{t_m}\right)^{-2/3}f_\Lambda(t_0, 0.16, - 0.01)\right]\right\}^{-2/\gamma}.
\end{aligned}
\end{equation}
\normalsize

As $t\rightarrow 0$, the mass of the halo becomes small compared to the final mass so that we can write,
\begin{equation}
\frac{M_h(t)}{10^{12}\mathrm{M}_\odot} \approx \left\{\left(\frac{M_0}{10^{12}\mathrm{M}_\odot}\right)^{-\gamma /2} + \frac{2 \gamma}{\sqrt{S_0} K_D t_m^{2}} \left(\frac{t}{t_m}\right)^{-2/3} \right\} ^{-2/\gamma},
\end{equation}
where the first term in the RHS is much smaller than the second term. This shows that masses of early haloes depend very weakly on their average final mass, and that the halo mass initially grows roughly $\propto t^{4}$, (since $\gamma\approx 1/3$).

Finally, collecting the numerical and cosmology dependent constants together, we can write \cref{A:eq:halo_mass} as,
\small
\begin{equation}
\begin{aligned}
\frac{M_h(t)}{10^{12}\mathrm{M}_\odot} & =  \\
& \left\{\left(\frac{M_0}{10^{12}\mathrm{M}_\odot}\right)^{-\gamma /2} + 0.31 \gamma \, 
    \left[\left(\frac{t}{t_m}\right)^{-2/3}f_\Lambda(t, 0.16, - 0.01) - 1.67 \right]\right\}^{-2/\gamma}
\end{aligned}
\end{equation}
\normalsize

\subsection{The halo mass function}

In the Press \& Schechter analysis, the co-moving abundance of haloes of mass $M_h$ at time $t$ is given by \citep{press_formation_1974},
\begin{equation}\label{A:eq:dndm_eps}
\frac{\dd n^{\textcolor{red}{c^*}}(M_h,t)}{\dd M_h} = \frac{{\rho}_0}{M_h^2} \frac{\delta_c \gamma}{\sqrt{2\pi} S^{1/2}} \frac{1}{D}
         \exp\left( - \frac{\delta_c^2}{2 S D^2}\right)  
\end{equation}
where we have assumed that the density power spectrum is a power law with exponent $\gamma$ and written the co-moving matter density of the Universe as ${\rho}_0$ following our convention. Using the evolution of the growth factor given by \cref{A:eq:D_of_t_approx_consts} and keeping the leading order terms we obtain,
\small
\begin{equation}\label{A:eq:halo_mass_fun}
    \begin{aligned}
        \frac{\dd n^{\textcolor{red}{c^*}}(M_h,t)}{\mathrm{dlog}_{10} M_h} & = \frac{2.94 \times 10^{-12}\,\mathrm{M}_\odot^{-1} \rho_0 \gamma}{\sqrt{S_0}K_D t_m^2}
         \, \left(\frac{M_h}{10^{12}\mathrm{M}_\odot}\right)^{-\left(1-\frac{\gamma}{2}\right)} \,\, \times \\
        &\left(\frac{t}{t_m}\right)^{-2/3} 
        f_\Lambda(t, 0.16, - 0.01) \,\,\times \\
         & \exp \left[ \frac{-5.14}{S_0 K_D^2 t_m^4}
         \left(\frac{M_h}{10^{12}\mathrm{M}_\odot}\right)^{\gamma}
         \left(\frac{t}{t_m}\right)^{-4/3} f_\Lambda(t,0.32, 0)\right].
    \end{aligned}
\end{equation}
\normalsize

For the cosmological parameters adopted in this paper, $\rho_0 = 3.913\times 10^{10} \Msol \Mpc^{-3}$. Substituting for values of the constants and cosmological parameters, we can write \cref{A:eq:halo_mass_fun} as,
\small
\begin{equation}
    \begin{aligned}
         \frac{\dd n(M_h,t)}{\mathrm{dlog}_{10} M_h} & = 5.43\times 10^{-3}
        \,\mathrm{cMpc}^{-3} \, \left(\frac{M_h}{10^{12}\mathrm{M}_\odot}\right)^{-\left(1-\frac{\gamma}{2}\right)} \,\, \times \\
        &\left(\frac{t}{t_m}\right)^{-2/3} 
        f_\Lambda(t, 0.16, - 0.01) \,\, \times \\ 
         & \exp \left[ -0.13 \left(\frac{M_h}{10^{12}\mathrm{M}_\odot}\right)^{\gamma}
         \left(\frac{t}{t_m}\right)^{-4/3} f_\Lambda(t,0.32, 0) \right].
    \end{aligned}
\end{equation}
\normalsize

\section{Comparison with hydrodynamical simulations}\label{sec:comparison}

\begin{figure}
\centering 
\includegraphics[width=0.48\textwidth]{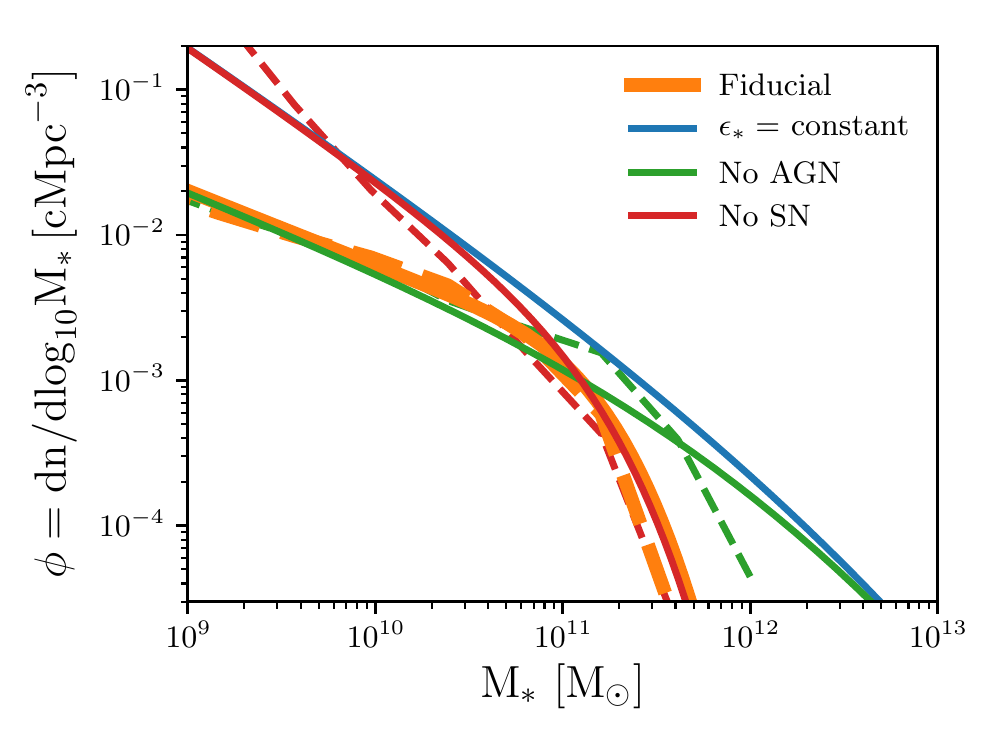}
  \vspace{-1.5em}
 \caption{Redshift $z=0.1$ GSMF for three variations of the \textsc{eagle} $(50\mathrm{cMpc})^{3}$ simulations at redshift $z{=}0$ (dashed lines), compared their equivalent analytic effective star formation efficiency model (solid lines). The orange line shows the Ref-L050N0752 \textsc{eagle} model. The ``No AGN'' and ``No SN'' models are shown in green and red respectively. While the models were not calibrated to reproduce their hydro simulation equivalent, they capture their overall behaviour reasonably well. The small differences are consistent with the differences in the efficiency parameters (see \cref{fig:SHMR_Eagle}).}
 \label{B:fig:GSMF}
\end{figure}

\begin{figure}
\centering 
\includegraphics[width=0.48\textwidth]{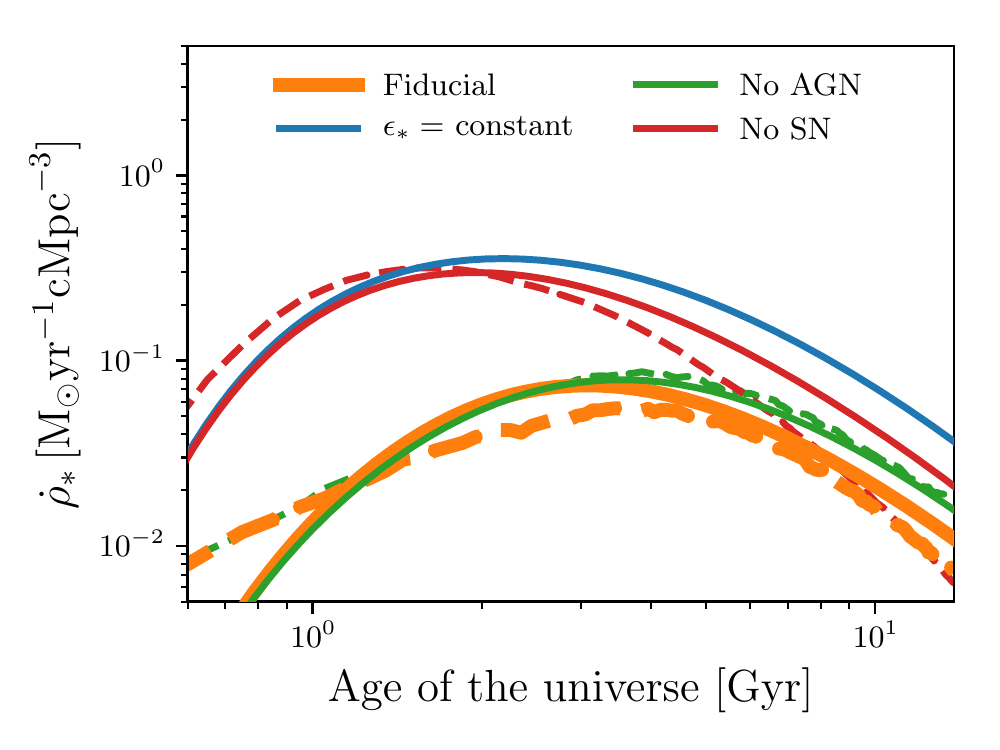}
  \vspace{-1.5em}
 \caption{SFR history of the Universe for three variations of the \textsc{eagle} $(50\mathrm{cMpc})^{3}$ simulations (dashed lines), compared their equivalent analytic effective star formation efficiency model (solid lines). The orange line shows the Ref-L050N0752 \textsc{eagle} model. The ``No AGN'' and ``No SN'' models are shown in green and red respectively. In order to compare with the simulations, we have included the effect of cosmic reionisation, $\epsilon_*(T_\mathrm{vir} < 10^4 \mathrm{K}) = 0$ in both the ``No SN'' and ``constant'' star formation efficiency models. While the models were not calibrated to reproduce their hydro simulation equivalent, they capture their overall behaviour reasonably well. The small differences are consistent with the differences in the efficiency parameters (see \cref{fig:SHMR_Eagle}).}
 \label{B:fig:rho_star}
\end{figure}

In this section, we compare three variations of the \textsc{eagle} $(50\mathrm{cMpc})^{3}$ simulations to their equivalent analytic effective star formation efficiency model. The Ref-L050N0752 \textsc{eagle} model \citep{schaye_eagle_2015,crain_eagle_2015,mcalpine_eagle_2016}, uses the same calibrated sub-grid parameters as the reference model $(100\mathrm{cMpc})^{3}$, ran with the same resolution, but in a smaller volume. The ``No AGN'' run uses the same calibrated sub-grid parameters as the reference model but removing feedback from BHs. For the ``No SN'' model \citep[red, introduced in][]{bower_dark_2017}, feedback from star formation has been removed. We note that, while the \textsc{eagle} ``No SN'' simulation removes the effect of star formation feedback, it still includes the effect of cosmic reionisation. Hence, there is a suppression of star formation is small haloes. In order to compare with the simulations, we have included the effect of cosmic reionisation, $\epsilon_*(T_\mathrm{vir} < 10^4 \mathrm{K}) = 0$ in both the ``No SN'' and ``constant'' star formation efficiency models. Finally, there is no \textsc{eagle} equivalent to the ``constant'' (or ``no feedback'') model. In \cref{B:fig:GSMF} we compare the GSMF at $z=0.1$, and in \cref{B:fig:rho_star}, we compare the SFR history of the Universe.

While much more computationally expensive, the behaviour of the full hydrodynamical simulations is well approximated by the analytic models introduced here.

%%%%%%%%%%%%%%%%%%%%%%%%%%%%%%%%%%%%%%%%%%%%%%%%%%

% Don't change these lines
\bsp	% typesetting comment
\label{lastpage}
\end{document}